\documentclass[reprint,longbibliography,
 amsmath,amssymb, aps]{revtex4-1}

\usepackage{graphicx}
\usepackage{dcolumn}
\usepackage{bm}
\usepackage{graphicx}
\usepackage{color}
\usepackage{bbold}

\usepackage{comment}
\usepackage[dvipsnames]{xcolor}
\usepackage[colorlinks=true,linkcolor=blue,citecolor=red,urlcolor=blue]{hyperref}
\newcommand{\bra}[1]{\langle{#1}|}
\newcommand{\ket}[1]{|{#1}\rangle}
\newcommand{\bkt}[2]{\langle{#1}|{#2}\rangle}

\usepackage{mathtools}
\usepackage[normalem]{ulem}

\def\sph{\mathcal{Y}}
\begin{document}

	\title{Green's function treatment of Rydberg molecules with spins}% 
	\author{Chris H. Greene$^1$ and Matthew T. Eiles$^2$}%
	% \email{chgreene@purdue.edu}
	
	\affiliation{
		$^1$Department of Physics and Astronomy and
		Purdue Quantum Science and Engineering Institute, Purdue University, West Lafayette, Indiana 47907, USA \\
		$^2$Max-Planck-Institut für Physik komplexer Systeme, Nöthnitzer Str. 38, 01187 Dresden, Germany\\
		\looseness=-3 % avoid linesbreaks in affiliations
	}

	\date{\today}% It is always \today, today,
	
	\begin{abstract}
		The determination of ultra-long-range molecular potential curves has been reformulated using the Coulomb Green's function to give a solution in terms of the roots of an analytical determinantal equation.  For a system consisting of one Rydberg atom with fine structure and a neutral perturbing ground state atom with hyperfine structure, the solution yields potential energy curves and wavefunctions in terms of the quantum defects of the Rydberg atom and the electron-perturber scattering phaseshifts and hyperfine splittings.  This method provides a promising alternative to the standard currently utilized method of diagonalization, which suffers from problematic convergence issues and nonuniqueness, and can potentially yield a more quantitative relationship between Rydberg molecule spectroscopy and electron-atom scattering phaseshifts.
	\end{abstract}
	
	\maketitle

	\section{Introduction\label{sec:intro}}
	For the past {fifteen} years, the spectroscopy of ultra-long-range Rydberg molecules consisting of a bond between one Rydberg atom and one or more ground state atoms has flourished and become increasingly quantitative \cite{bendkowskyObservation2009,bendkowskyRydberg2010,gajMolecular2014,sassmannshausenExperimental2015,boothProduction2015a,schlagmullerProbing2016a,niederprumRydberg2016,niederprumObservation2016,camargoCreation2018a,maclennanDeeply2019,whalenProbing2019,engelPrecision2019,whalenFormation2019,dingCreation2020,deissObservation2020,whalenHeteronuclear2020,baiCesium2020a,kanungoLoss2020,peperPhotodissociation2020a,peperHeteronuclear2021,peperRole2023}. The basic picture of the bonding mechanism remains rooted in the scattering of the Rydberg electron by the ground state perturbing atom, which is qualitatively the same as in early studies of Rydberg-neutral interactions \cite{duInteraction1987,DuGreene1987errat,duMultichannel1989,greeneCreation2000a,hamiltonShaperesonanceinduced2002}.  This picture produces unusual oscillatory Born-Oppenheimer potential curves and intriguing electron probability distributions for some of the electronic states, which resemble trilobites or butterflies \cite{greeneCreation2000a,boothProduction2015a,hamiltonShaperesonanceinduced2002,niederprumObservation2016}.  The aim of this present study is to implement a  Green's function treatment that includes the effects of all spin interactions in order to make first principles theory far more quantitative than existing theoretical techniques.
	
	Since the earliest experimental observations of this class of molecular bound states, the original picture of the atom-atom bonding has been confirmed in its basics.\cite{greeneCreation2000a} That picture derives from the Fermi-Omont representation of the effective energy associated with the electron-atom scattering phaseshifts $\delta_L(\epsilon)$, but while it has been confirmed, its limitations have also become apparent.  The Fermi-Omont effective zero-range interaction terms in the Hamiltonian \cite{fermiSopra1934,omontTheory1977} are proportional to $\tan{\delta_L(\epsilon)}$ where $\epsilon$ is the kinetic energy of the electron at the point where it collides with the perturbing electron.  For systems possessing a low energy shape resonance, like the e-Rb and e-Cs systems, that tangent function can diverge to infinity. This is the case in particular for the butterfly Rydberg molecule states whose potential curves are controlled by a $^3P^o$ scattering resonance \cite{chibisovEnergies2002a,hamiltonShaperesonanceinduced2002}.  The resulting divergence poses a stringent challenge for any theory aiming to quantitatively describe the molecular spectroscopy.  
	
	The presence of occasionally divergent terms in the Hamiltonian produces instabilities that require renormalization when standard methods for computing energy eigenvalues are utilized, such as diagonalization of $H$ in a truncated expansion into an orthonormal basis set.  That  diagonalization approach has until now been the method of choice for calculations of the Rydberg molecule Born-Oppenheimer potential energy curves; this preference is in part because of the relative ease with which one can add additional spin-spin and spin-orbit interaction terms to the Hamiltonian, when higher precision is desired \cite{eilesHamiltonian2017,eilesTrilobites2019,AndersonPRA,eilesFormation2018,Markson}. (While Ref.~\cite{Markson} and Ref.~\cite{eilesHamiltonian2017} have both developed Hamiltonians for the diagonalization method which include all of the spin degrees of freedom treated in the present study, we recommend that calculations using this method should preferably implement the final result of Ref.~\cite{eilesHamiltonian2017}, for the reasons discussed in that article.) A drawback is that the expansion is known to not converge, leading to non-unique basis size-dependent potentials that make it challenging, if not impossible, to compare experiment and theory in a fully objective manner \cite{feyComparative2015}.
	Furthermore, the electron energy needed to evaluate the energy-dependent scattering phase shifts cannot be determined self-consistently in such diagonalization approaches, leading to a lack of self-consistency \cite{peperHeteronuclear2021}. 
	
	Our treatment implements a nonperturbative  Green's function description of the electronic energy eigenstates of such Rydberg-ground state diatomic molecules, which incorporates in principle all spin-dependent interactions in addition to the basic Coulombic Hamiltonian. The goal is to make fully quantitative the mapping of electron-atom scattering information and the Rydberg atomic quantum defects into accurate Born-Oppenheimer potential energy curves whose rovibrational states can be measured and tested accurately.  The  Green's function treatment developed here has been motivated by an earlier Kirchoff integral formulation \cite{chibisovEnergies2002a,hamiltonShaperesonanceinduced2002,khuskivadzeAdiabatic2002}, which does not suffer from non-convergence issues that plague numerical diagonalization treatments. Because the method manipulates the {\it phases} of the wavefunction near the perturbing atom instead of performing a diagonalization of the Hamiltonian based on the scattering volume-dependent and energy-dependent pseudopotentials, the divergences of $\tan{\delta}$ cause no difficulties in the Green's function theory. Our resulting extension of that Kirchoff integral Green's function method to include all spin dependent interactions, presented here, is expected to significantly reduce much of the current uncertainty in comparing theoretical and experimental Rydberg molecule energy levels.  With this improvement, it should allow sharper conclusions to be drawn in deducing electron-atom scattering information from spectroscopic measurements. Other promising approaches to the calculation of Rydberg molecule spectra have been developed in recent years~\cite{taranaLongrange2020,Giannakeas2020a}, but they have not yet been extended to include the full set of spin-orbit and hyperfine interactions built into the present treatment.
	\section{\label{sec:level1}Theory}
	\subsection{The basic ideas and notation}

 \begin{table*}
   \label{tab:qnumbers}
\caption{Summary of the operators and other quantities used here to describe long-range Rydberg dimers involving two alkali atoms. The  hyperfine details assume that the perturber atom is $^{87}$Rb. The zero of our energy scale is the statistical average of perturber hyperfine energy levels plus the positive ion ground state energy.}
  \begin{tabular}{ c | c | c }
    \hline
    \textbf{Operator} & \textbf{eigenvalues} & \textbf{Physical meaning} \\ \hline
    $\vec L$ & $L=0,1$ & electron orbital angular momentum relative to the perturber \\ \hline
$\vec \ell$ & $\ell=0,1\dots n-1$ & electron orbital angular momentum relative to the Rydberg core \\ \hline 
$\vec s_R$ & $s_r=\frac{1}{2}$ & Rydberg electron spin \\ \hline
$\vec s_p$ & $s_p=\frac{1}{2}$ & perturber electron spin \\ \hline
$\vec I$ & $I=\frac{3}{2}$ & perturber nuclear spin\\ \hline
$\vec f$ & $f=1,2$ & total angular momentum of the perturber \\ \hline
$\vec S$ & $S=0,1$ & total electron spin of the molecule \\ \hline
$\vec j$ & $j=\frac{1}{2},\frac{3}{2},\dots n-\frac{1}{2}$ & total angular momentum of the Rydberg electron relative to its core \\ \hline
$\vec J$ & $J=0,1,2$ & total electronic angular momentum of the anion, relative to its core \\ \hline
\hline
\textbf{Quantity} & \textbf{Possible values} & \textbf{Physical meaning}
 \\\hline
 $m_R,m_p$ & $\pm \frac{1}{2}$ & electronic  spin magnetic quantum numbers\\\hline
 $m_I$ & $\pm \frac{3}{2},\pm \frac{1}{2}$ & nuclear  spin magnetic quantum number\\\hline
 $m_\ell=m_L$ & $|m_\ell|\le 1$ & projection of the Rydberg electron orbital angular momentum onto $\hat z$\\\hline
 $M_\text{tot}=m_\ell + m_R + m_p + m_I$ & $|M_\text{tot}|\le \frac{7}{2}$ & projection of the total angular momentum onto $\hat z$\\\hline
 $\mu_{\ell,j}$ & - & quantum defect of the Rydberg atom \\ \hline
  $\delta^{SLJ}$ & - & scattering phase shift of the electron-perturber\\\hline
  $\nu$ & $\nu =(-2E)^{-1/2}$ & the effective principal quantum number defined at $E<0$\\\hline
  $n$ & $n=1,2\dots$ & integer-valued principal quantum number for hydrogenic Rydberg states\\\hline
  \end{tabular}
    
\end{table*}
	We use atomic units throughout, based on the reduced mass of the atomic ion - electron system. 
	The full Hilbert space of interest includes the Rydberg electron's position and orbital angular momentum operators,  $\vec{r}$ and $\vec{\ell}$ respectively, relative to the Rydberg core. Relative to the perturber, at a position $\vec{R}=R{\hat z}$, these operators are denoted  $\vec{X}=\vec{r}-\vec{R}$ and $\vec{L}$. 
	The remainder of the state space of interest includes multiple spin operators: the Rydberg electron spin $\vec{s}_R$, the perturber atom electronic spin $\vec{s}_p$, and the perturber atom nuclear spin $\vec I$. The nuclear spin $I$ should not be confused with the identity operator, $\bf 1$. The Rydberg molecules studied experimentally to date have involved perturbing ground state atoms from either the first or second column of the periodic table, which have no orbital angular momentum. Our formulas omit reference to the Rydberg atom's nuclear spin quantum number $I_R$ since its associated hyperfine structure decreases rapidly with $\ell$ and with $n$ as $n^{-3}$, and is typically negligible for currently achievable spectroscopic resolution. However, if ever appropriate, this can be readily incorporated without changing the basic structure of our approach. 

	In addition to these operators, some key intermediate angular momentum sum operators used in the following are the total angular momentum operator of the perturber,  $\vec{f}=\vec{s}_p+\vec{I}$, and the total electronic spin of the molecule,
	$\vec{S}=\vec{s}_R+\vec{s}_p$.  Other summed angular momenta that enter our treatment include the total angular momentum of the Rydberg electron relative to the positive ion nucleus,
	$\vec{j}=\vec{s}_R+\vec{\ell}$, and the total electronic angular momentum of the electron-perturber system relative to the perturbing atom, namely $\vec{J}=\vec{L}+\vec{S}$.
	The following formulation does not explicitly describe the Rydberg electron as ``indistinguishable'' from the perturber electron(s), although their indistinguishability is understood to have been incorporated when computing the scattering phaseshifts of a free electron (i.e. the Rydberg electron in the present context) from the perturbing atom.
	Accordingly, a complete set of states in the Hilbert space is the ket
	\begin{equation}
		|r,\ell,m_\ell;s_R,m_{R};s_p,m_{s_p};I,m_{I}\rangle,
		\label{eq:ketdef}
	\end{equation}
	where we imply that this state has an angular dependence associated with a spherical harmonic $\sph_{l,m_l}(\theta,\phi)$ with respect to the Rydberg core. 
Note that we use $\sph$ to denote spherical harmonics in order to avoid confusion with the variable $Y$.
Table~\ref{tab:qnumbers} summarizes this list of different quantities and their allowed values. 
	
	Space is partitioned into three parts: a small sphere centered on the Rydberg core, where short-ranged interactions between the multielectron core and the Rydberg electron produce quantum defects $\mu_{\ell,j}$, a second small sphere centered on the neutral perturber, where the Rydberg electron scatters with known phase shifts $\delta^{S,L,J}(k)$, and finally, the vast region of space outside of these two spheres. In this latter region the Hamiltonian is written
	\begin{align}
		\label{eq:hamdef}
		H_0=&h_{\rm Ryd}\otimes{\bf 1}_{s_p}\otimes{\bf 1}_{I}
		\\&+ {\bf 1}_{\vec{r}}\otimes{\bf 1}_{s_R}\otimes\sum_{f,m_{f}}\ket{ fm_{f} }E_{f} \bra{ fm_{f}}.\nonumber
	\end{align}
	Here $h_{\rm Ryd}$ is the full short range Hamiltonian of the Rydberg electron in the field of the positive ion nucleus and any screening or spectator core electrons in the singly-charged ionic ground state.
 The unit operators for various degrees of freedom are also indicated in this Hamiltonian in an obvious notation.
 The state $\ket{fm_{f}} \equiv \ket{ (s_p I)fm_{f} }$ is the coupled nuclear spin state of the perturber, and its hyperfine energy levels $E_{f}$ are given by
	\begin{equation}
		E_{f}=\frac{A}{2}[f(f+1)-I(I+1)-s_p(s_p+1)]. 
	\end{equation}
	$A$ is the hyperfine structure constant of the perturber. 
 The Rydberg electron energy levels $E_{n\ell j}$ are determined by the single channel quantum defects via
	\begin{equation}
		\label{eq:rydenergy}
		E_{n\ell j}=-\frac{1}{2(n-\mu_{\ell j})^2}.
	\end{equation}
	This can be generalized to a multichannel quantum defect theory (MQDT) representation of the atomic Rydberg levels \cite{Seaton,LuFano,FanoJOSA}, as in Ref.~\cite{Eiles2015}, when appropriate. 
	
	If electron-perturber interactions are neglected, and if quantization of the radial $\{r\}$ degree of freedom is initially postponed in the spirit of MQDT, the set of channels are characterized by the commuting observables for $H_0$. This set of channel functions has the structure
	\begin{equation}
		\label{eq:channeldef}
		|i\rangle \equiv |(s_R \ell)jm_{j},fm_{f}\rangle, 
	\end{equation}
	corresponding to channel threshold energies equal to
	\begin{equation}
		\label{eq:channelenergy}
		E_i \equiv E_{f}.
	\end{equation}
	This representation describes a set of diagonal potentials that, in the absence of electron-perturber interactions, have no coupling whatsoever:
	\begin{equation}
		\label{eq:channelpotentials}
		V_{ii'}(r)=\delta_{ii'} \biggr( E_{i}-\frac{1}{r}+\frac{\ell(\ell+1)}{2r^2} \biggr),
	\end{equation}
for electron-ion distances $r>r_0$.  Here it is assumed that there is a relatively small distance $r_0 \sim 20$ a.u.. beyond which the electron-ion interaction potential can be approximated as purely Coulombic.  Of course at distances $r<r_0$ the Rydberg electron experiences complex screening and exchange interactions with the ionic core electrons, but the effect of those complex interactions is fully encapsulated in the weakly energy-dependent quantum defects $\mu_{\ell j}$.  
	
	 Green's function methods can be implemented in various alternative approaches.  For instance, one can impose physical boundary conditions at $r\rightarrow \infty$ at the outset, or one can postpone the large-$r$ finiteness boundary condition in the spirit of MQDT, which would imply utilizing the so-called ``smooth  Green's function'',  $G^{(S)}$ \cite{Fano1978,StrinatiGreene,duInteraction1987,DuGreene1987errat}.  Our present formulation utilizes the former method based on the physical Green's function.  Owing to the spin degrees of freedom in this system, the Green's function $G$ for $H_0$ is a diagonal matrix in the spin indices, which at total energy $E$ satisfies
 \begin{align}
		\label{eq:greenfuncdef}
		(H_0-E)G^{C}({\vec r},{\vec r\,'})&=\delta({\vec r}-{\vec r}\,')\delta_{m_{R}fm_{f},m_{R}'f'm_f'}.
	\end{align}
	In particular, the present treatment adopts the Green's function that imposes the correct boundary conditions for the Rydberg electron as it emerges from the ionic core in a specified $(\ell s_R)j$ channel with its appropriate $j$-dependent quantum defect.  This is especially important for the Rydberg $p$ and $d$ states. 

	\subsection{Review of the key Green's function equations}
	The method developed here builds on the original spin-independent formulation of Hamilton \cite{hamiltonPhotoionization}, which in turn is based on the closed-form Coulomb Green's function derived by Hostler and Pratt \cite{HostlerPratt}.
	This Green's function, which ignores spin degrees of freedom and atomic quantum defects, is
	\begin{align}
		\label{eq:hostlergreenfunction}
		G^{C}(\vec{r},\vec{r} \, ' \! ;\nu) =
		\frac{\Gamma(1-\nu)}{2 \pi |\vec{r}-\vec{r} \, '|}
		\Bigg[& M'_{\nu,1/2}(\eta) W_{\nu,1/2}(\xi) \\\nonumber&-
		M_{\nu,1/2}(\eta) W'_{\nu,1/2}(\xi)
		\Bigg],
		\label{eq:GFanal}
	\end{align}
	where $W$ and $M$ are standard Whittaker functions, primes denote ordinary derivatives with respect to the argument, $\nu = \sqrt{-2E}$ is the principal quantum number defined at $E<0$, and 
	\begin{equation}
		\label{eq:arguments}
		\xi,\eta = (r + r' \pm |\vec{r}-\vec{r} \,' |)/\nu.
	\end{equation}
	Quantum defects can be included naturally via a correction term to be added to $G^C$, derived originally by Davydkin {\it et al.} \cite{Davydkin1971},
	\begin{widetext}
		\begin{align}
			\label{eq:greenfunctionqdef}
			G^{q.d.}(\vec{r},\vec{r} \, ' \! ;\nu) &= \sum_{\ell =0}^{\ell_0}\sum_{m_\ell=-\ell}^{m_\ell=\ell}
			\lambda_{\ell}(\nu)\sph_{\ell m_\ell}(\theta,\phi) \sph_{\ell m_\ell}^{*}(\theta',\phi') {\cal W}_{\nu,\ell +1/2}(r)
			{\cal W}_{\nu,\ell +1/2}(r')/(rr'),
		\end{align}
  \end{widetext}
		where $\ell_0$ is the highest angular momentum with non-vanishing quantum defect and
		\begin{equation}
			\label{eq:lambda}
			\lambda_{\ell}(\nu)\equiv \frac{\pi \sin{\pi \mu_{\ell }}}{\sin{\pi \nu} \sin{\pi(\mu_{\ell }+\nu)}}.
		\end{equation}
		The rescaled Whittaker functions in Eq. \ref{eq:greenfunctionqdef},
		\begin{equation}
			\label{eq:whittaker}
			{\cal{W}}_{\nu,l+1/2}(r)=\sqrt{\frac{\nu}{\Gamma (\nu-\ell) \Gamma (\ell+\nu+1)}} W_{\nu,\ell+\frac{1}{2}}\left(\frac{2 r}{\nu}\right),
		\end{equation} have an ``energy-normalized'' amplitude $\sqrt{2} (\pi k(r))^{-1/2}$ in the classically allowed region.\cite{Aymar1996} 
	
		A first key generalization needed here in order to incorporate the full spin-dependent physics of the Rydberg molecule is to include the perturber's hyperfine structure as a constant energy shift dependent on the $f$ quantum number. 
		This is done by inserting the identity operator $\sum_{f,m_f}\ket{fm_f}\bra{fm_f}$ into Eq.~\ref{eq:hostlergreenfunction} and evaluating the Green's function at the principal quantum number $\nu_f$, where $E-E_f \equiv -\frac{1}{2{\nu_f}^2}$.
		Eq. \ref{eq:hostlergreenfunction} becomes
		\begin{align}
			\label{eq:greenfunchyperfine}
			{\hat G}^C(\vec{r},\vec{r}';\nu)&=\sum_{m_R,f,m_f} |s_Rm_R,f m_f\rangle G^C_{\nu_f}(\vec{r},\vec{r}\,') \langle f m_f,s_Rm_R|,
		\end{align}
		which is also diagonal with respect to the Rydberg electron's spin.
		The generalization of Eq. \ref{eq:greenfunctionqdef} includes the $j$-dependent quantum defects of the Rydberg electron, yielding
  \begin{widetext}
		\begin{align}
			\label{eq:greenfunchyperfinefull}
			{\hat G}^{q.d.}(\vec{r},\vec{r} \, ' \! ;\nu) &=
			\sum_{f,m_f} \sum_{\ell,j,m_j}
			\lambda_{\ell ,j}(\nu_f)\frac{ {\cal W}_{\nu_f,\ell +1/2}(r)
				{\cal W}_{\nu_f,\ell +1/2}(r')}{rr'} 
			\times\ket{j m_j}\bra{j m_j}\times |fm_f\rangle \langle fm_f |.
		\end{align}
	\end{widetext}
	The quantity $\lambda_{\ell,j}(\nu_f)$ is identical to $\lambda_{\ell}(\nu)$, but generalized to accept $j$-dependent quantum defects, $\mu_\ell\to\mu_{\ell,j}$.
	The full operator is the sum,
	${\hat G}= {\hat G}^C+{\hat G}^{q.d.}. $
	In order to organize the angular momentum and spin indices, we introduce a shorthand notation:

 \begin{align}
		\label{eq:ishort}
		{\bf i} &\equiv {L,M_L, m_R, f, m_f},
	\end{align}
 which defines also a corresponding state
 \begin{equation}
       \ket{\bf i} \equiv \ket{s_Rm_R,(s_pI)fm_f,LM_L}.
 \end{equation}
	Since the Coulomb Green's function of Eq. \ref{eq:greenfunchyperfine} is diagonal in all of the ${\bf i}$ quantum numbers {\it except} for $L$, another useful index will be ${\bf\bar{i}}$, consisting of all ${\bf i}$ quantum numbers except $L$.  
 We define $\delta_{{\bf \bar{i},\bar{i}'}} \equiv \delta_{f,f'}\delta_{m_{f},m_{f'}}\delta_{m_R,m_R'}\delta_{M_{L},M_{L}'}$. 
 Note that the size of the set of quantum numbers included in ${\bf i} $, which will eventually set the dimension of the determinantal equation, is restricted by the number of partial waves included, the hyperfine spin of the perturber, and the projection of the total angular momentum onto the internuclear axis, $M_\text{tot}$, which is a good quantum number. 
 The number of quantum defects included does not affect this size, nor does the energy of the Rydberg states in question.
 This is an important difference between this approach and diagonalization of a basis expansion, where the matrix dimension grows linearly with $\nu$.

	\subsection{Green's function terms evaluated near the perturber}
 	Now that the Green's function matrix has been determined, the electron-perturber scattering information will be incorporated into the calculation simply as boundary information on the tiny sphere centered on the perturber. 
	To set the stage for this, this section presents analytic results for the Green's function matrix in the vicinity of the perturber. 
	The  vectors $\vec X$ and $\vec Y$ are defined according to
	$
	\vec{r}\equiv \vec{R}+\vec{X}$ and 
	$\vec{r} \,' \equiv \vec{R}+\vec{Y}$, with the goal in mind that all relevant Green's functions or Rydberg wave functions will be evaluated at these coordinates. 
	The next step is to evaluate all of these expressions at $Y \ll X \ll 1$, corresponding to a tiny region around the perturber. 
	The Green's operator in the ${\bf i}$ representation can be expressed conveniently in these coordinates.
 
 The Green's function expansion about the perturber position reads 
 \begin{equation}
     G_{\nu_f}^C(\vec r,\vec r)\approx \sum_{L,L'}\sum_{M_L}\sph_{LM_L}(\hat X)\Delta_{L,L'}^{M_L}(X,Y)\sph^*_{L'M_L}(\hat Y).\nonumber
 \end{equation}
	Analytical expressions for $\Delta_{L,L'}^{M_L}(X,Y)$, to lowest  order in $X$ and $Y$, are given in Appendix \ref{app:taylor}, specifically Eqs. \ref{eq:del} and \ref{eq:dell}.
 These terms vanish when $M_L \ne M_L'$ due to cylindrical symmetry. 
	Insertion of this expansion into Eq. \ref{eq:greenfunchyperfine}
	yields
	\begin{align}
		\label{eq:greenfunchyperfine1}
		{\hat G}^C(\vec{r},\vec{r}';\nu)&=\bra{\hat X}\left[\sum_{{\bf i},{\bf i'}}\ket{\bf i}\delta_{{\bf\bar{i}},{\bf\bar{i'}}} \Delta^{M_L}_{L,L'}(X,Y) \bra{\bf i'}\right]\ket{\hat Y}.
	\end{align}
The number of partial waves in the expansion about the perturber has been restricted here to $L\le 1$, which is generally sufficient for long-range Rydberg molecules. We will discuss the possible effects of higher partial waves \cite{Giannakeas2020b} in more detail in Section \ref{sec:discussion}. 
Should it become necessary to extend the present treatment to higher partial waves, the derivations in Appendix~\ref{app:taylor} can be generalized.

	To obtain a similar form as Eq. \ref{eq:greenfunchyperfine1} for the quantum defect correction in Eq. \ref{eq:greenfunchyperfinefull},  the Taylor expansion of the Coulomb wave function is needed for small $X$ and $Y$. 
	These expansions are derived in Appendix \ref{app:taylorwhit}, namely Eq. \ref{eq:b00}-\ref{eq:b11}, and are
	\begin{equation}
		\label{eq:taylorexp}
		\frac{ {\cal W}_{\nu_f,\ell+1/2}(r)}{r} Y^{(\ell)}_{M_L}({\hat r}) \approx \sum_{L=|M_L|}^1 X^L b^{(\ell,\nu_f)}_{L M_L} Y^{(L)}_{M_L}({\hat X}).
	\end{equation}
	We insert this expansion, the analogous expansion of the $r'$ dependence in terms of the variable $Y$, and the same identity operators in terms of $L$ and $M_L$ as above, into Eq. \ref{eq:greenfunchyperfinefull}, obtaining
	\begin{widetext}
		\begin{align}
			\label{eq:qdcorr2}
			G^{q.d.}(\vec r,\vec r\,';\nu)= \bra{\hat X}\left\{\sum_{{\bf i},{\bf i'}}\ket{{\bf i}} \delta_{f,f'}\delta_{m_{f},m_{f'}} X^{L}Y^{L'} \sum_{\ell,j,m_j} \lambda_{\ell,j}(\nu_f)b^{(\ell,\nu_f)}_{L, M_{L}}\left[ b^{(\ell,\nu_f)}_{L', M_{L'}}\right]^*{\cal S}^{\ell,j}_{M_{L},M_{L'},m_{R},m_{R'}}\bra{\bf i'}\right\}\ket{\hat Y},
		\end{align}
  \end{widetext}
		where 
		\begin{align}
			{\cal S}^{\ell,j}_{M_{L},M_{L'},m_{R},m_{R'}}&\equiv \sum_{m_j} C_{lM_{L},s_Rm_{R}}^{jm_j}C_{lM_{L'},s_Rm_{R'}}^{jm_j}.
		\end{align}
		 Eqs. \ref{eq:greenfunchyperfine1} and \ref{eq:qdcorr2} give a useful form for the full Green's function when it will be evaluated in the vicinity of the perturber, at small $X$ and small $Y$:
		\begin{equation}
			\label{eq:gfmatrixelements}
			G(\vec r, \vec r\,';\nu)=\sum_{{\bf i},{\bf i'}}\bkt{\hat X}{{\bf i}}\left(G^C_{{\bf i},{\bf i'}}(X,Y)+G^{q.d.}_{{\bf i},{\bf i'}}(X,Y)\right)\bkt{{\bf i'}}{\hat Y}.
		\end{equation} 
	
	\subsection{Integral equation}
	\label{sec:integral}
	The next step utilizes this spin-dependent Coulomb Green's function treatment to calculate the molecular potential energy curves.  
	The derivation starts from the following integral equation for the electronic wave function $\Psi(\vec r)$, valid everywhere outside of a small volume of radius $Y$ around the perturber:
		\begin{align}
			 \Psi(\vec{r}) = \frac{Y^2}{2}  \oint \Bigg\{
			&\frac{\partial G(\vec{r},\vec{r} \, ')}{\partial Y}
			\Psi(\vec{r} \, ')
			-\frac{\partial \Psi(\vec{r} \, ')}{\partial Y}
			G(\vec{r},\vec{r} \, ') 
			\Bigg\} d\hat Y .
			\label{eq:OuterSolution3}
		\end{align}
	The most convenient expression for the  Rydberg electron's wave function in the perturber's vicinity is its partial wave expansion, since a small collection of energy-dependent spin-orbit coupled phase shifts $\delta^{SLJ}(k)$ suffice to parameterize the full wave function. 
	As with the shorthand index ${\bf i}$, here it is useful to define a second index, 
	
 \begin{align}
		\label{eq:ashort}
		{\boldsymbol \alpha} &\equiv {S,L,J,M_J,I,m_{I}},
	\end{align}
 along with the state 
 \begin{equation}
   \ket{\boldsymbol{\alpha}} \equiv \ket{[(s_Rs_p)SL]JM_J,Im_I},
 \end{equation}

	incorporating all of the degrees of freedom of the perturber spins and atom-electron scattering complex. 
 The size of the set ${\bf \alpha}$ is equal to that of ${\bf i}$.  
	In the $\ket{\boldsymbol\alpha}$ representation the wave function near the perturber is
	\begin{align}
		\label{eq:wfnearperturber1}
		\bkt{\vec r}{\Psi} &= \sum_{\boldsymbol\alpha} B_{\boldsymbol\alpha}\Phi_{\boldsymbol\alpha}(k,Y)\bkt{\hat Y}{\boldsymbol\alpha}.
	\end{align}
		For $Y$ sufficiently large that the perturber-electron potential has vanished, but small enough that the Coulomb potential is effectively constant, the
		radial wave function is given in terms of spherical Bessel functions $j_L$ and $y_L$,
		\begin{align}
			\label{eq:wfnearperturber2}
			\Phi_{\boldsymbol\alpha}(k,Y)&= j_L(k Y) \cos{\delta^{SLJ}(k)}-y_L(k Y) \sin{\delta^{SLJ}(k)}.
		\end{align}
		Some discussion is needed of the meaning of Eqs \ref{eq:wfnearperturber2}, and in particular, the choice of the electron momentum $k$.  When hyperfine structure can be ignored, $k$ is obtained semiclassically via $k=\sqrt{-\frac{1}{\nu^2}+\frac{2}{R}}$. This expression already makes one approximation, namely that the electron-ion reduced mass is equal to the electron-perturber reduced mass, certainly adequately accurate in typical applications.  (Recall that our choice of units throughout this article is atomic units based on the electron-ion reduced mass, set here to unity.)  A further complication arises when the perturber possesses very low-lying energy levels such as hyperfine structure, because when the Rydberg electron collides with the perturber atom with kinetic energy equal semiclassically to $\varepsilon = -\frac{1}{2 \nu^2} +\frac{1}{R}$, it is actually a multichannel problem and there is an electron-perturber scattering matrix with different wavenumbers $k_f$ in the different hyperfine channels. For this situation, we interpret the rest of our derivation in this article, based on Eq.~\ref{eq:wfnearperturber}, as making a frame transformation approximation, as in the spirit of Refs.~\cite{RauFano, Lee1975, Greene1987}, this will normally be an excellent approximation because $k_f \approx k$, but it can begin to fail (implying a need for improvement) in those limited ranges of $R$ where $k\rightarrow 0$.
		
		Inserting the expression for the Green's function near the perturber (Eq.~\ref{eq:gfmatrixelements}) and the wave function near the perturber (Eq. \ref{eq:wfnearperturber2})  into Eq.~\ref{eq:OuterSolution3} yields
  \begin{widetext}
\begin{align}
			\label{eq:wfnearperturber}
			\sum_{\boldsymbol{\alpha}}\Phi_{\boldsymbol\alpha}(k,X)\bkt{\hat X}{\boldsymbol\alpha}B_{\boldsymbol\alpha} = \frac{1}{2} \sum_{{\bf i,i'},\boldsymbol\alpha} \bkt{\hat X}{{\bf i}}
			Y^2 \oint\left\{\left[\partial_Y G_{\bf i,i'}(X,Y) \Phi_{\boldsymbol\alpha}(k,Y) 
			- G_{\bf i,i'}(X,Y)  \partial_Y \Phi_{\boldsymbol\alpha}(k,Y)\right] \langle{\bf i'}\ket{\hat Y}\bra{\hat Y}\boldsymbol\alpha \rangle\right\}d{\hat Y} B_{\boldsymbol\alpha}.    
		\end{align}
		Because of the form of $G_{\bf i,i'}(X,Y)$ derived in Eq. \ref{eq:gfmatrixelements}, the integration over $\hat Y$ is trivially removed by by the presence of the identity operator in Eq. \ref{eq:wfnearperturber}. 
		The transformation of the integral equation in Eq. \ref{eq:wfnearperturber1} into a matrix equation is accomplished now by projecting onto $\bra{\boldsymbol\alpha'}$ and $\oint d\hat X\ket{\hat X}$ from the left:
		\begin{align}
			\label{eq:gfstep}
			\Phi_{\boldsymbol{\alpha'}}(k,X) B_{\boldsymbol{\alpha'}}
			= \frac{1}{2} \sum_{{\bf i,i'},\boldsymbol{\alpha}}   \bkt {\boldsymbol{\alpha'}}{{\bf i}}
			Y^2 \Bigg[\partial_Y G_{\bf i,i'}(X,Y)  \Phi_{\boldsymbol\alpha}(k,Y)- G_{\bf i,i'}(X,Y)  \partial_Y \Phi_{\boldsymbol\alpha}(k,Y)\Bigg]\bkt{ {\bf i'}}{\boldsymbol\alpha}B_{\boldsymbol{\alpha}}  .
		\end{align}
	The spin recoupling matrix elements $\langle {\bf i}|\boldsymbol\alpha \rangle$ are
	\begin{align}
		\label{eq:recoup}
		\mathcal{A}_{i\alpha}\equiv \langle{\bf i}  | \boldsymbol{\alpha}  \rangle =\langle \boldsymbol{\alpha} | {\bf i} \rangle &=  
		\delta_{{L_i},{L_\alpha}}\sum_{M_{S},{m_{p}}}
		C_{S M_{S},L_\alpha M_{L}}^{J M_{J}}  C_{s_R m_{R},s_pm_{p}}^{S M_{S}} C_{s_p m_{p},I m_{I}}^{f m_{f}}.
	\end{align}
	Eq. \ref{eq:gfstep} is the key equation, but it is not yet ready to implement, until we expand everything to lowest orders in $X$ and $Y$, again assuming that $Y\ll X$ throughout. 
	This has already been accomplished for the Green's functions, using the equations Eqs. \ref{eq:del}, \ref{eq:dell}, and \ref{eq:qdcorr2}, and additionally the expansion of 
	\begin{equation}
		\bar{\Delta}^{M_L}_{L,L'}(X,Y)  = \oint\text{d}\hat X\oint \text{d}\hat Y \sph_{LM_L}^*(\hat X)  \partial_Y\label{eq:aafb} G^C_\nu(\vec{X},\vec{Y})\sph_{L'M_L}(\hat Y)
	\end{equation}
	derived in Eq. \ref{eq:delbar} in Appendix \ref{app:taylor}. 
	After plugging in the lowest terms in these expansions and those for the spherical Bessel functions in the radial wave function $\Phi_{\boldsymbol{\alpha}}(k,Y)$, we finally obtain the set of linear equations to be solved for the electronic energies. 
	With the definition of $\bar{B}_\alpha \equiv B_\alpha (\frac{k}{3})^{L_\alpha}$, the final key set of linear equations becomes:

  \begin{equation}
		\label{eq:deteq}
		\sum_{\alpha} \left(-\delta_{\alpha'\alpha} \cos{\delta^{S_\alpha,L_\alpha,J_\alpha}} 
			+\Omega_{\alpha'\alpha}\right)\bar{B}_\alpha\equiv \sum_\alpha M_{\alpha'\alpha}(\nu,R) \bar{B}_\alpha=0,
	\end{equation}
	in terms of the matrix  
		\begin{align}
  \label{eq:omegadefinition}
			\Omega_{\alpha'\alpha}&=\frac{\sin \delta^{S_\alpha,L_\alpha,J_\alpha}}{k^{2L_\alpha+1}}\sum_{i,i'}\mathcal{A}^T_{\alpha' i}\left[\sum_{K'=0}^{1}\sum_{K=0}^{1}\delta_{L_{\alpha'},K'}\delta_{L_\alpha,K}\left(\delta_{{\bf\bar{i}},{\bf\bar{i}'}}P_{K'+1,K+1}(\nu_{f_i},R)+\frac{8L_\alpha+1}{2}Q_{ii'}\right)\right]\mathcal{A}_{i'\alpha}\
		\end{align}
		where
		\begin{align}
  \label{eq:qdefinition}
			Q_{ii'}&\equiv \delta_{f,f'}\delta_{m_{f},m_{f}'}\sum_{\ell,j} \lambda_{\ell,j}(\nu_f)b^{(\ell,\nu_{f'})}_{L_i, M_{L}} b^{(\ell,\nu_{f})}_{{L_i}', M_{L}'}   {\cal S}^{\ell,j}_{M_{L},M_{L}',m_{R},m_{R}'}.
		\end{align}
	\end{widetext}
	
	Here $b^{(\ell,\nu_{f})}_{L_i, M_{L}}$ is defined in Appendix \ref{app:taylorwhit}, Eqs.~\ref{eq:b00},~\ref{eq:b10}, and~\ref{eq:b11},  and with $K \in \{0,1\}$, the $P_{K'K}$ are matrix elements of
	\begin{align}
 \label{eq:pdefinition}
		P &= \begin{pmatrix}
			\Phi_v &  3\sqrt{3}\Phi_{uv}\\
			\frac{1}{\sqrt{3}}\Phi_{uv} & -\Phi_{vvv}+3\delta_{M_{L}',0}\Phi_{uuv}
		\end{pmatrix}.
	\end{align}

	\subsection{Numerical procedure}
	Equation \ref{eq:deteq} is satisfied only at discrete values of $\nu$, which we obtain numerically by finding the roots of $\det \mathbf{M}(\nu,R)$ at each $R$. 
    In practice, to make the root-finding procedure more stable, we conduct the numerical search for roots in the quantity
    \begin{align}
      \nonumber  \tilde{\mathbf{M}}(\nu,R)=&\mathrm{sign}[\Pi_{l,j,f}\sin\pi(\mu_{l,j} + \nu_f)\det \mathbf{M}(\nu,R)]\\&\times \left|\Pi_{l,j,f}\sin\pi(\mu_{l,j} + \nu_f)\det \mathbf{M}(\nu,R)\right|^{1/7}. 
      \label{eq:detactual}
    \end{align}
		 The product $\Pi_{l,j,f}$ is taken over all $l$ and $j$ values with non-zero quantum defects and over all hyperfine levels; this term removes singularities stemming from $\lambda_{l,j}(\nu_f)$. 
   The choice of a power of $1/7$ is to ``smooth" the variation (of many orders of magnitude) in the determinant as a function of $\nu$, and the numerical procedure could be improved in different regimes by adjusting this. 

   The unusual characteristics of Rydberg molecule potential curves makes it challenging in some cases to obtain these roots. 
   The breaking of different symmetries by the relativistic or hyperfine couplings in the Hamiltonian is often quite weak, which means that the oscillatory potential curves can frequently become nearly degenerate, impeding the resolution of multiple roots. 
   Furthermore, these potential curves can change from a rather smooth variation in regions where the electronic state is primarily in a Rydberg state of low angular momentum, to rapid variations when the electronic state becomes dominated by high angular momentum trilobite and butterfly states. 
   This combination of rapid fluctuations and near degeneracies complicates the choice of search grid used to numerically find the roots. 

   To resolve these issues, in a first pass the potential energy curves are computedfor a large value of $M_\text{tot}$ where the number of roots, and in particular of nearly degenerate roots, is diminished. 
   We exploit the adiabaticity of the potential curves by starting our search at large $R$, where the threshold values of the potential curves are known, and proceeding inwards to small $R$ by searching for each discrete root $\nu(R_i)$ within a series of energy windows bracketing the roots found at $R_{i+1}$. 
	For smaller $M_\text{tot}$ values, we implement this same process, but also include search windows centered around the roots found for higher $M_\text{tot}$. 
 This helps to treat regions where nearly degenerate potential energy curves vary rapidly as a function of $R$.

It is interesting that the computation of highly excited spectra of quantum billiards is often accomplished by solving a very similar determinantal equation also derived using the relevant Green's function
 \cite{backerNumerical2002}. 
	Some approximation methods developed in this context may be useful here \cite{vebleExpanded2007}. 
Alternatively, it may be advantageous to search instead for zeroes in the eigenvalues of $\mathbf{M}(\nu,R)$ in order to avoid missing roots due to near-degeneracies. 

\begin{figure}[t]
  \includegraphics[width=0.95\columnwidth]{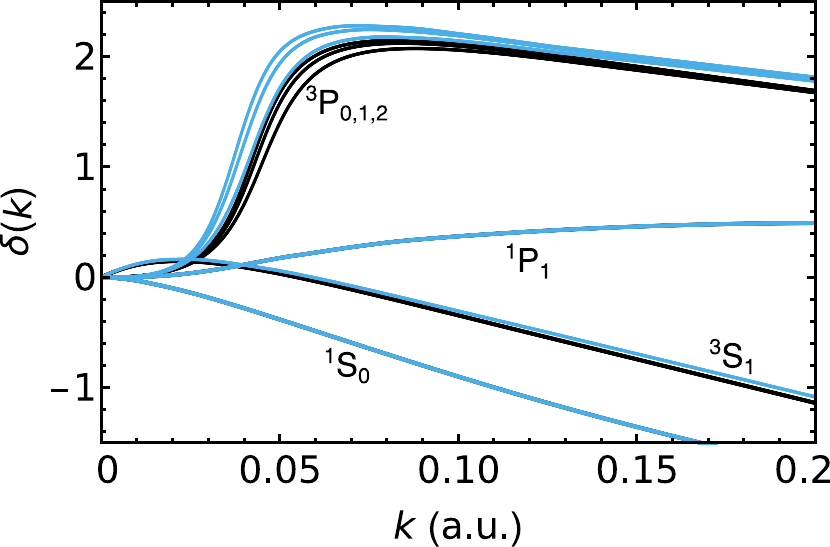}
		\caption{\label{fig:phases}  
Electron-Rb scattering phase shifts used in this paper. The black (dark) curves are  calculated \textit{ab initio} using a relativistic model potential \cite{khuskivadzeAdiabatic2002}. The blue (light) curves show phase shifts which were modified (in the triplet case only) by Ref.~\cite{engelPrecision2019} to produce electronic potential curves whose vibrational energies match those observed in experiment .  	}
	\end{figure}

 \subsection{Atomic parameters}
 
 The hyperfine splitting, atomic quantum defects, and polarizabilities utilized in our calculations have been determined via precision spectroscopy to very high accuracy and are collected in Ref. \cite{eilesTrilobites2019}. 
 The electron-atom scattering phase shifts, on the other hand, are only available from theoretical calculations and the values of key properties, such as the zero-energy scattering lengths and shape resonance widths and positions,  vary from source to source \cite{eilesFormation2018,eilesTrilobites2019}. 
 For example, relativistic e-Rb phase shifts for $L\le 1$ were published by Fabrikant and coworkers in Ref.~\cite{khuskivadzeAdiabatic2002}, and are shown as black(dark) curves in  Fig. \ref{fig:phases}.
The blue(light) curves in this figure are phase shifts which were fit to experimental data taken in $s$-state Rydberg molecules \cite{engelPrecision2019}. 
 This was accomplished by varying the triplet scattering phases so that the potential energy curves - computed using a basis set benchmarked to the results of a spin-independent Green's function calculation - predicted vibrational states and binding energies in agreement with the measurements \cite{engelPrecision2019}. 
 The zero-energy scattering length obtained from this calculation is about 10\% smaller than the \textit{ab initio} value, but its value remains within the spread of scattering lengths obtained theoretically and from similar experiments \cite{eilesTrilobites2019,maclennanDeeply2019,bendkowskyRydberg2010}. 
 The position of the $P$-wave resonance is about 20\% smaller than calculated.

The next section presents potential curves obtained from our present Green's function method, alongside those obtained by diagonalizing the zero-range Hamiltonian. These illustrate that our present method can be reliably used to obtain accurate potential energy curves without the debilitating dependence on the basis size. 
From these comparisons, we emphasize that the phase shifts previously obtained from the diagonalization method are most likely valid only as model-dependent fit parameters. 
This is due to the inability to benchmark the potential curves with alternative methods, and worse, the fact that the dependence of the potential curves on the input phaseshifts varies with internuclear distance, and measurements taken over a finite range of Rydberg levels or bond lengths likely do not provide unique fits. 
We will further emphasize this point by comparing the potential curves obtained from the Green's function method with these two sets of phase shifts, showing that the current ``best set" of phase shifts are incompatible with experiment when used outside the scope of the model. 

 \begin{figure}[t]
		\includegraphics[width=0.95\columnwidth]{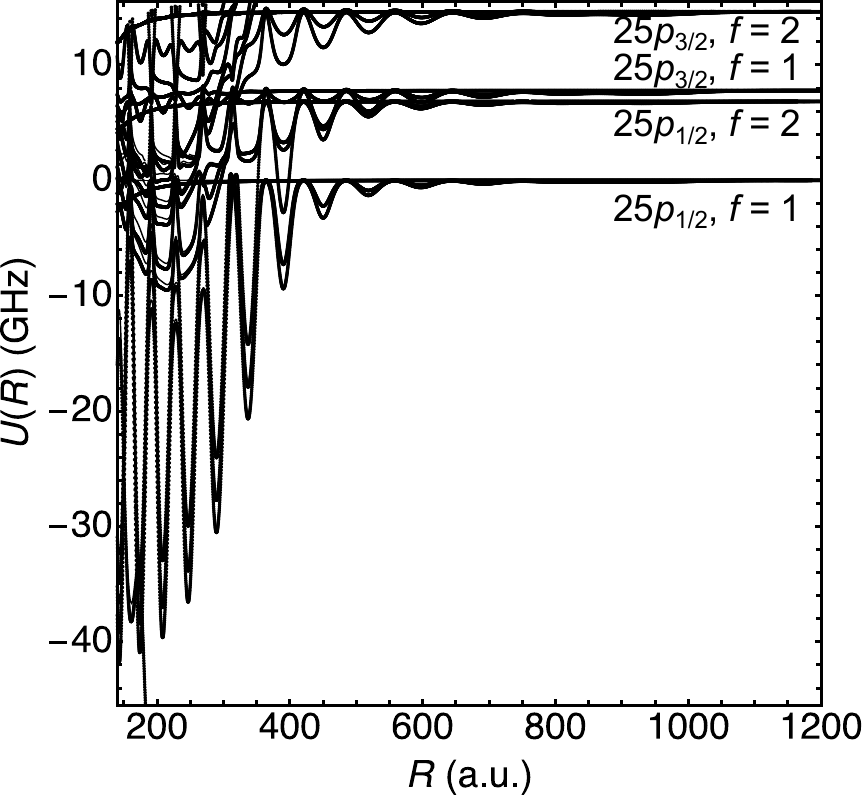}
		\caption{\label{fig:n23all}  
Potential energy curves with threshold values at the $25p_j,f=1,2$ levels, presented relative to the $25p_{1/2},f=1$ threshold. The deep wells at smaller $R$ values host butterfly molecular states. The calculation uses the fitted phase shifts of Ref.~\cite{engelPrecision2019}. 
		}
	\end{figure}
	\section{Results}
   After they are found numerically, the collection of roots $\nu_i(R)$ determine the potential energy curves 
  \begin{equation}
  \label{eq:potencurves}
      U_i(R) = -\frac{1}{2\nu_i(R)^2}-\frac{\alpha}{2R^4},
  \end{equation}
  where the second term denotes the polarization interaction between the Rydberg ion and the neutral atom. 

   \begin{figure}[t]
		\includegraphics[width=0.95\columnwidth]{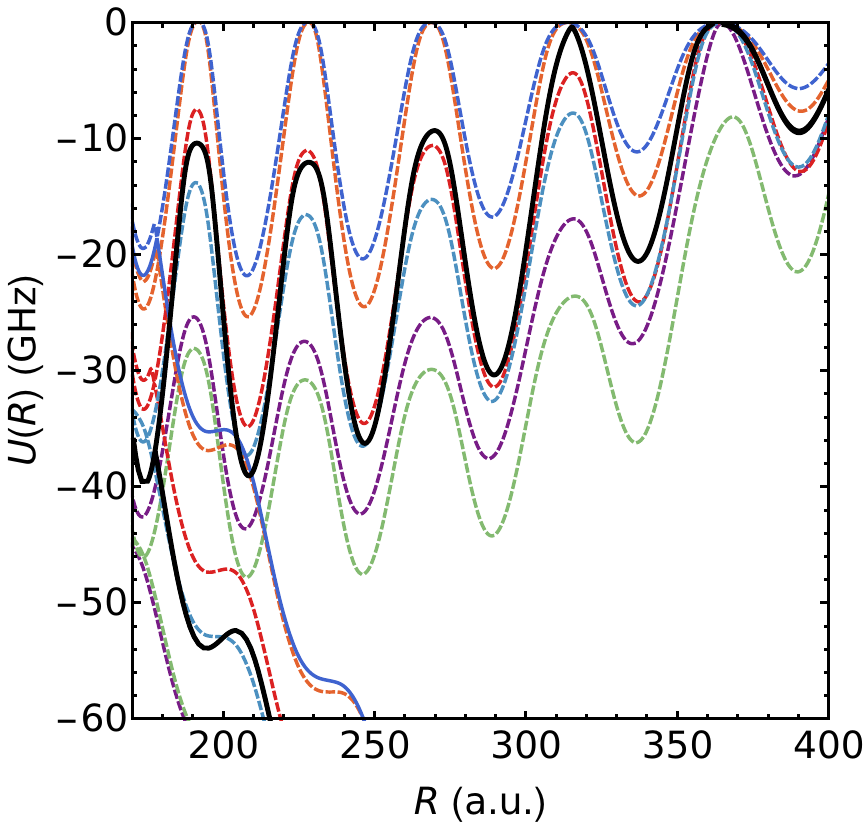}
		\caption{\label{fig:n23groundstates}
The deepest butterfly potential energy curve near the $25p_{1/2},f=1$ state representing the zero of the energy scale, calculated using our Green's function method (in black) and compared with diagonalization in 6 different basis sizes (colored and dashed). The basis sets include states from the degenerate Rydberg manifolds $22\le n \le 23$ (purple),  $21\le n \le 23$ (blue), $21\le n \le 24$ (cyan), $21\le n \le 25$ (green), $20\le n \le 24$ (orange), and $20\le n \le 25$ (red), as well as all quantum defect-shifted states within these energy ranges. These curves show the potential curves due to electron-atom scattering only, i.e. without the additional polarization interaction from Eq.~\ref{eq:potencurves}.
		}
	\end{figure}
	\subsection{Rydberg \textit{p} states and butterfly molecules}

	A fertile environment for testing our method is provided by the molecular states asymptotically reaching the unperturbed $(n+2)p_j$ Rydberg states of rubidium. 
 The label $n$ here represents the hydrogenic manifold lying energetically above these states, which are shifted by the large quantum defects of Rb. 
  The relevant potential energy curves, with energies displayed relative to that of the $25p_{1/2},f=1$ asymptote, are shown in Fig. \ref{fig:n23all}. 
 At large $R$, the molecular electronic character resembles that of the unperturbed atomic state, and the potential curves are subsequently of the shallow ``low-$\ell$" type. 
 Ref.~\cite{niederprumRydberg2016} reports
 molecular spectroscopy in this energy range.
 At smaller $R$ (around 400 a.u., the butterfly states descending from the unperturbed manifold of degenerate high-$\ell$ states with principal quantum number $n$ mix strongly with the $(n+2)p_j$ states. 
 This interaction dramatically deepens the potential wells. 
 The deepest vibrational states, having bond lengths between 150 and 350 a.u., were observed in Ref.~\cite{niederprumObservation2016}. 
 The measurement of large electric dipole moments in these molecular states confirmed that they were butterfly states. 

The relative depths of the butterfly molecule potential wells, at $R<400$a.u., are sensitive to the position of the $P$-wave shape resonance, the energy dependence of the $P$-wave phase shifts above resonance, and, due to the divergence of the $P$-wave term in the zero-range pseudopotential, the number of included Rydberg basis states in the diagoanlization method. 
We illustrate this in Fig. \ref{fig:n23groundstates} by plotting the deepest butterfly potential curve computed with various basis sets and comparing those with the present Green's function method.

   \begin{figure}[t]
		\includegraphics[width=0.95\columnwidth]{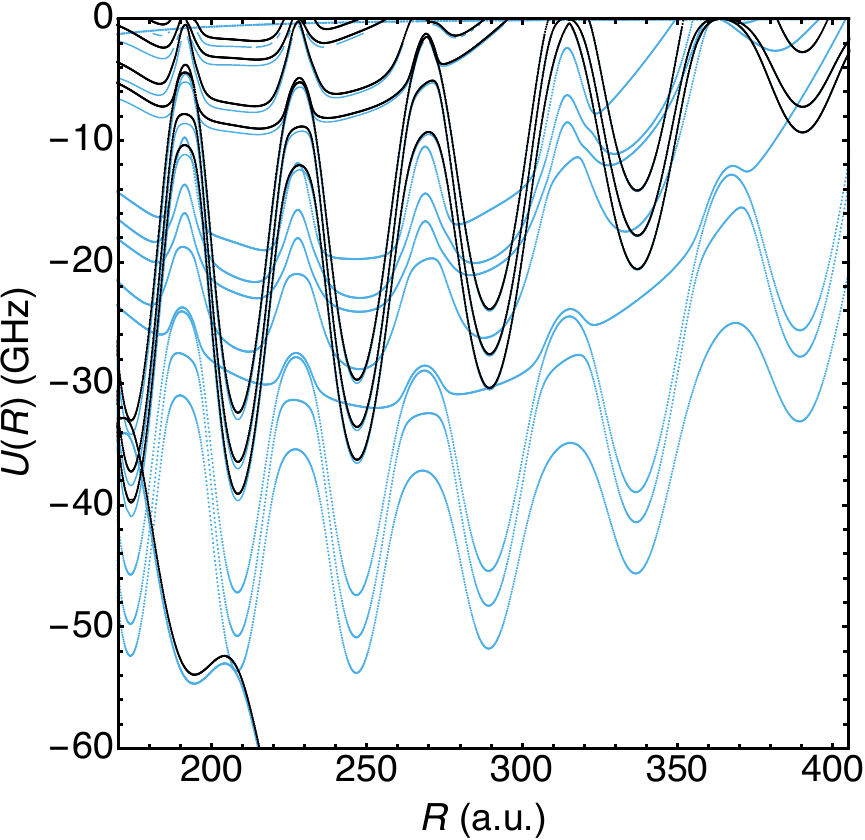}
		\caption{\label{fig:n23PhaseComparison}  Potential energy curves for Rb$^*$Rb supporting butterfly molecules with $M_\text{tot}=\frac{1}{2}$. At large $R$ these potential curves approach threshold values at the non-interacting $25p_j,f=2$ energy levels. The black curves were calculated with the fitted phase shifts of Ref.~\cite{engelPrecision2019}, and those in blue using the calculated phase shifts of Ref.~\cite{khuskivadzeAdiabatic2002}. Vibrational levels and their dipole moments were reported in Ref.~\cite{niederprumObservation2016} in the energy range from $-50$ GHz to $-40$ GHz. 
		}
	\end{figure}

 The calculated curves using the two basis sizes which were used in recent papers, Ref. \cite{engelPrecision2019} and \cite{deissObservation2020}, are shown in purple and orange. 
 Both of these disagree with the Green's function results by, on average, around 10 GHz; moreover, the diagonalization results do not converge to the Green's function result as the basis increases. 
 The agreement between a given diagonalization calculation and the Green's function also varies with internuclear distance, calling into question the validity of fitting basis set sizes to specific vibrational states in one or two potential wells alone.

 		\begin{figure}[t]
		\includegraphics[width=0.95\columnwidth]{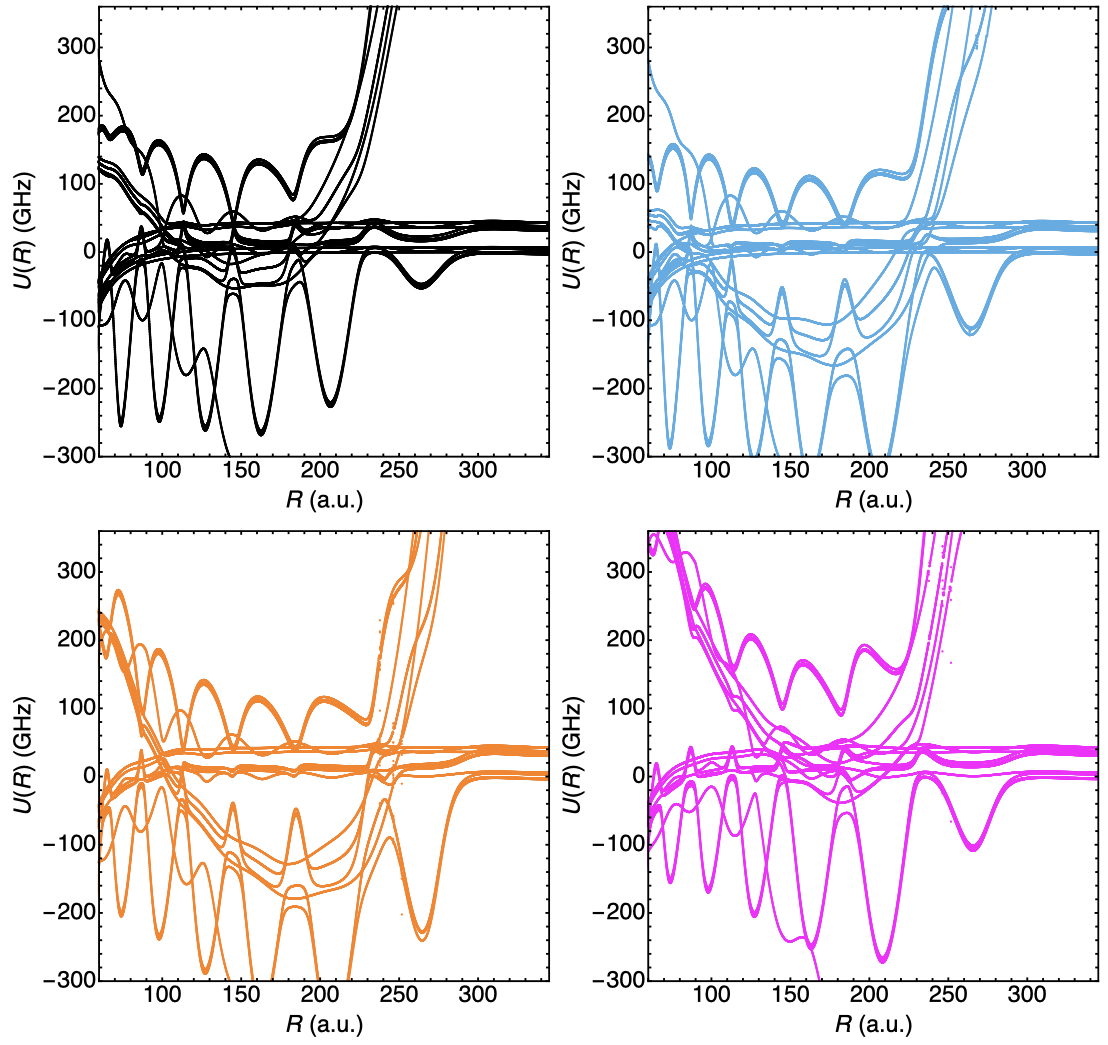}
		\caption{\label{fig:n14butterflies}  
Potential energy curves for Rb$^*$Rb with threshold values at the $16p_j,f=1,2$ levels, presented relative to the $16p_{1/2},f=1$ threshold. The upper left (black) Green's function calculation uses the fitted phase shifts of Ref.~\cite{engelPrecision2019}, while the upper right (blue) Green's function calculation uses the calculated phase shifts of Ref.~\cite{khuskivadzeAdiabatic2002}.
The bottom two panels show results using the fitted phase shifts and obtained via diagonalization with two different basis sizes. In the bottom left (orange) the basis includes states with $12\le n \le 14$; in the bottom right (magenta) the basis include $12 \le n \le 15$
}
	\end{figure}

 		\begin{figure}[t]
		\includegraphics[width=0.9\columnwidth]{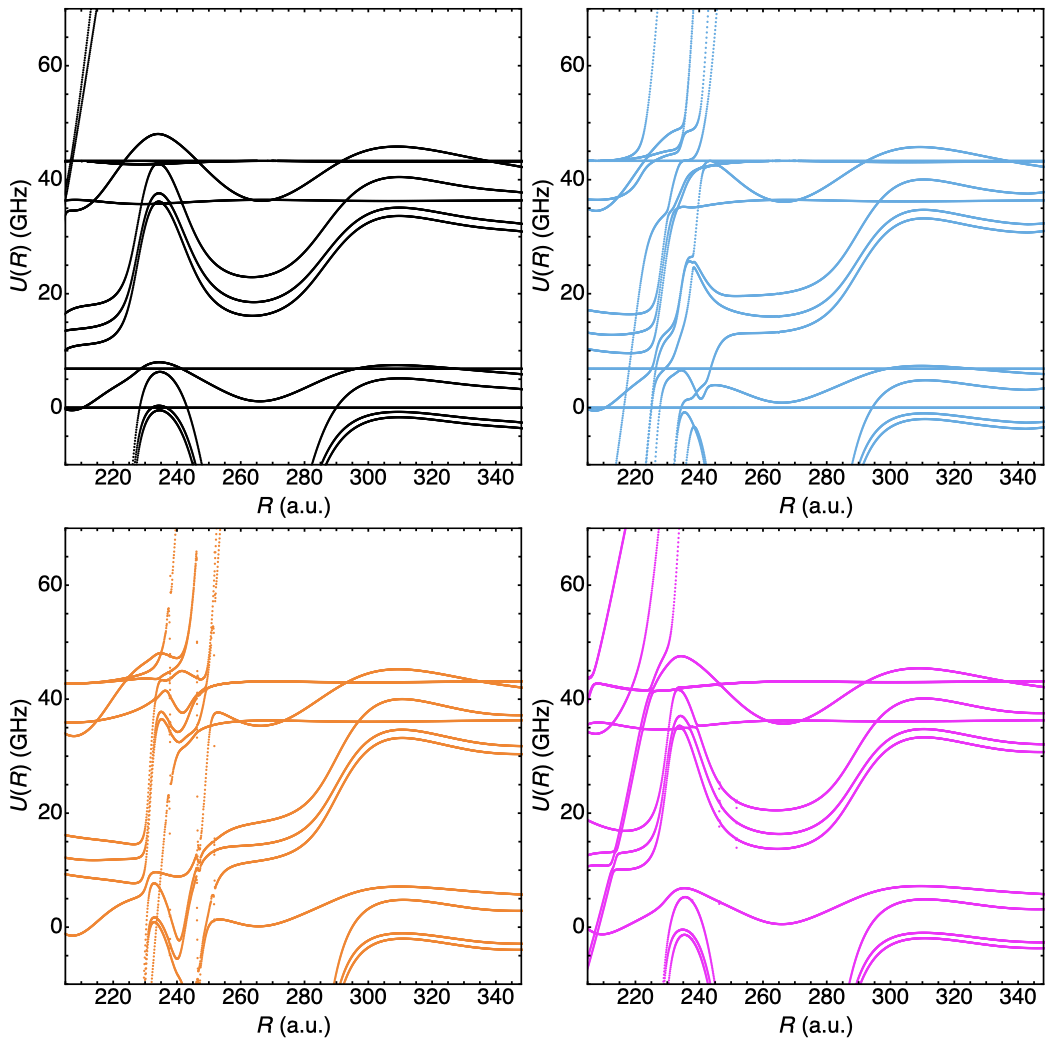}
	\caption{\label{fig:n14butterfliesZ}  
Potential energy curves for Rb$^*$Rb with threshold values at the $16p_j,f=1,2$ levels, presented relative to the $16p_{1/2},f=1$ threshold. These panels show the potential wells supporting the vibrational levels detected in Ref.~\cite{deissObservation2020}. The upper left (black) Green's function calculation uses the fitted phase shifts of Ref.~\cite{engelPrecision2019}, while the upper right (blue) Green's function calculation uses the calculated phase shifts of Ref.~\cite{khuskivadzeAdiabatic2002}.
The bottom two panels show results using the fitted phase shifts and obtained via diagonalization with two different basis sizes. In the bottom left (orange) the basis includes states with $12\le n \le 14$; in the bottom right (magenta) the basis include $12 \le n \le 15$
}
	\end{figure}

 Fig. \ref{fig:n23PhaseComparison} explores this further by comparing Green's function PECs calculated using the fitted phase shifts of Ref.~\cite{engelPrecision2019} (in black) and the calculated \textit{ab initio} phase shifts of Ref.~\cite{khuskivadzeAdiabatic2002} (in blue). 
The results obtained from the modified scattering phase shifts are 15-20 GHz shallower than those calculated from the calculated phase shifts over the full range of $R$. 
The modified phase shifts are therefore inconsistent with the experimental measurements of the butterfly-type vibrational bound states of Ref. \cite{niederprumObservation2016}, which were detected for energies above -50GHz. 
We conclude that the modified phase shifts of Ref.~\cite{engelPrecision2019}
remain model-dependent fitting parameters which allowed for accurate reproduction of the $ns$-state molecules detected there but are not the correct, model-independent, scattering parameters.

However, one should not prematurely conclude from Fig.~\ref{fig:n23PhaseComparison} that the calculated phase shifts are correct, even though they provide qualitatively better calculations than the fitted phase shifts for these butterfly states. We illustrate this using the same class of $(n+2)p$ states in Figs.~\ref{fig:n14butterflies} and \ref{fig:n14butterfliesZ}, where now $n=14$. 
The potential curves in the upper subfigures are again comparing the two different sets of phase shifts using the Green's function method, while the bottom two figures show the potentials obtained from diagonalization using the fitted phase shifts.  
The interesting region here are the three potential wells highlighted in Fig.~\ref{fig:n14butterfliesZ}, which were found to support vibrational levels via molecular spectroscopy reported in Ref.~\cite{deissObservation2020}. Here, the perturbation on the $(n+2)p$ states due to the butterfly potentials occurs at a larger $R$ for the calculated phase shifts than for the fitted phase shifts, disrupting the inner wall of this potential well. 
This would dramatically change, if not eliminate, the positions of vibrational lines in this well, in contradiction to the experimental evidence. 
It is clear from the bottom two panels why attempts to fit phase shifts with the diagonalization is not effective; the differences due to non-convergence in the basis set size can be patched over by modifying the phase shifts. 
Future work should re-investigate these comparisons between experiment and theory for these $16p$ molecular states. 
A new fit of the phase shifts using the method presented here will likely show that the physical phase shifts are in between these two sets compared here.

	\begin{figure}[t]
		\includegraphics[width=\columnwidth]{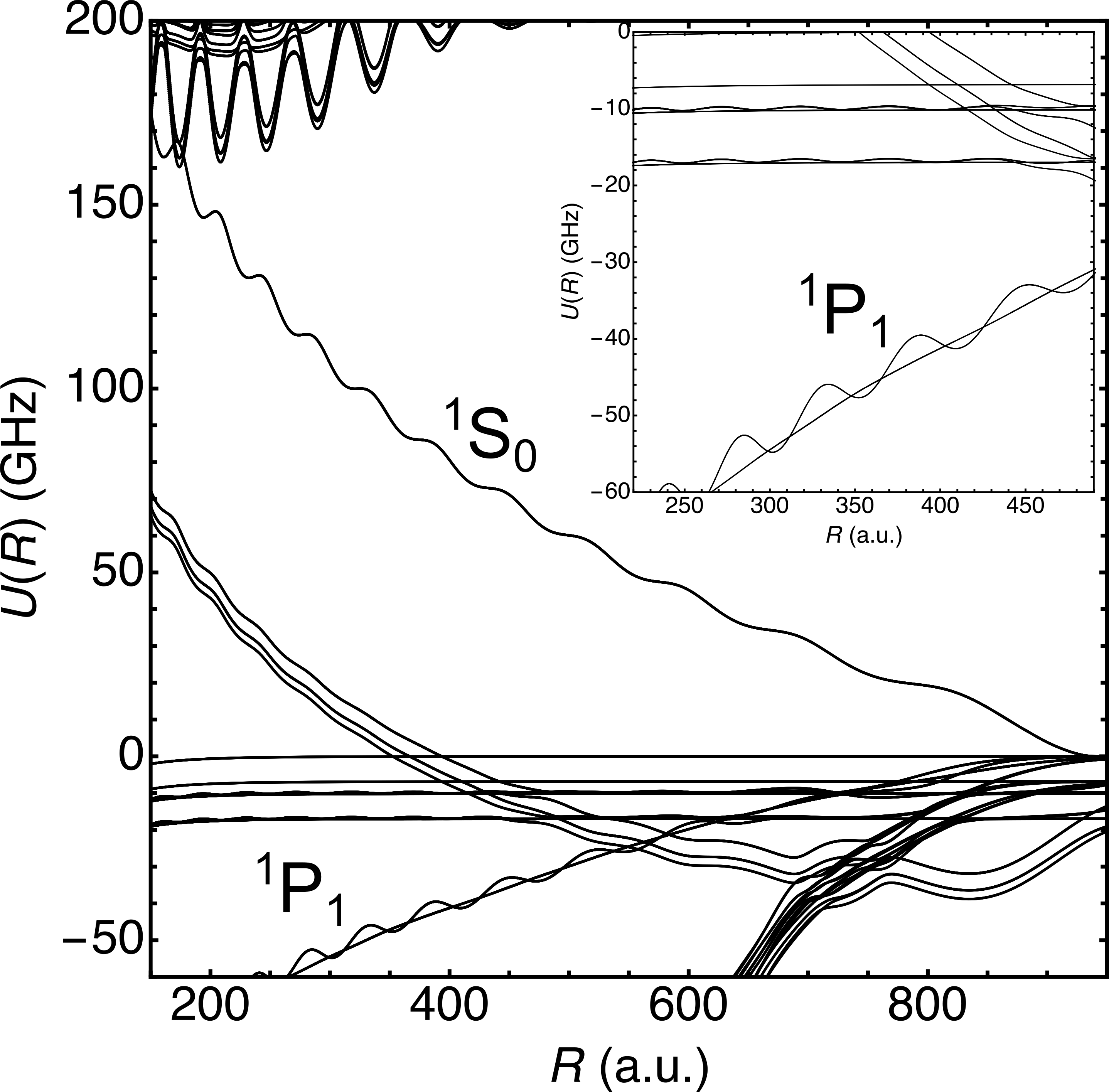}
		\caption{\label{fig:singlettotriplet23}  Potential energy curves of the singlet-dominated trilobite and singlet-dominated butterfly for Rb$^*$Rb and $M_\text{tot}=\frac{1}{2}$. The energies are measured relative to the hydrogenic level with $n=22$ and a perturber in the $f = 2$ state. The inset shows the three different singlet butterfly potential energy curves. One is oscillatory and two, indistinguishable on this scale, are smooth. 
		}
	\end{figure}

\subsection{Effect of hyperfine splitting across the Rydberg series}
Although the previous section demonstrated the quantitative differences between the present method and diagonalization of the zero-range pseudopotential, both methods yield qualitatively the same results. 
Exploration of new types of molecular features can in principle be done with either method equally well, but having a complementary approach at hand can, through its relative benefits or disadvantages, lead to new realizations. 
In this section we discuss two previously unnoticed effects related to the interplay between the different energy scales - the hyperfine splitting, the splitting between singlet and triplet energies, and Rydberg level density - in Rydberg molecules.

\begin{figure}[t]
  \includegraphics[width=\columnwidth]{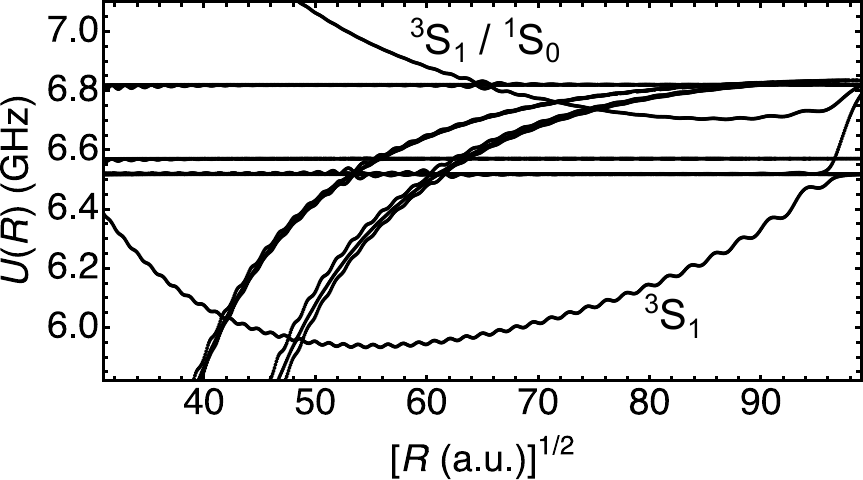}
		\caption{\label{fig:singlettotriplet70}  Trilobite potential energy curves for Rb$^*$Rb with threshold energy at the hydrogenic $n=70$ level with an $f=2$ perturber, with $M_\text{tot}=\frac{1}{2}$. The reference energy is set to the $f=1$ level. The lower trilobite level remains a triplet state, while the upper one is a mixture of singlet and triplet states. 
		}
	\end{figure}

Fig.~\ref{fig:singlettotriplet23} shows the potential energy curves with $M_\text{tot}=\frac{1}{2}$ in the vicinity of the degenerate manifold of hydrogenic states with $n=22$, just below the $25p$ potential curves discussed in the previous section. 
At this relatively low principal quantum number, the hyperfine splitting is small compared the energy shifts caused by the perturber on the hydrogenic states, and the total electron spin of the perturber-electron complex can be considered to be an approximately good quantum number for the trilobite and butterfly potential curves. 
The singlet trilobite potential curve is everywhere positive due to the monotonically increasing positive scattering length in the singlet channel. 
Contrast this with the same trilobite level at a much higher energy, as shown in Fig.~\ref{fig:singlettotriplet70}. 
Here, for $n=70$, the hyperfine splitting is much larger than the energy scale of the potential energy curves, and couples singlet and triplet scattering states together. The upper potential curve develops a large well due to its strong admixture of triplet character. 
Such a well could support bound states localized at very large internuclear distances, in contrast to the very broad triplet-dominated trilobite potential below. 
Note that this calculation does not include quantum defects for $\ell >4$ in order to highlight this well; a more careful study including the finite quantum defects of high-$\ell$ states would be necessary to quantitatively investigate its properties. 

\begin{figure}[b]
		\includegraphics[width=\columnwidth]{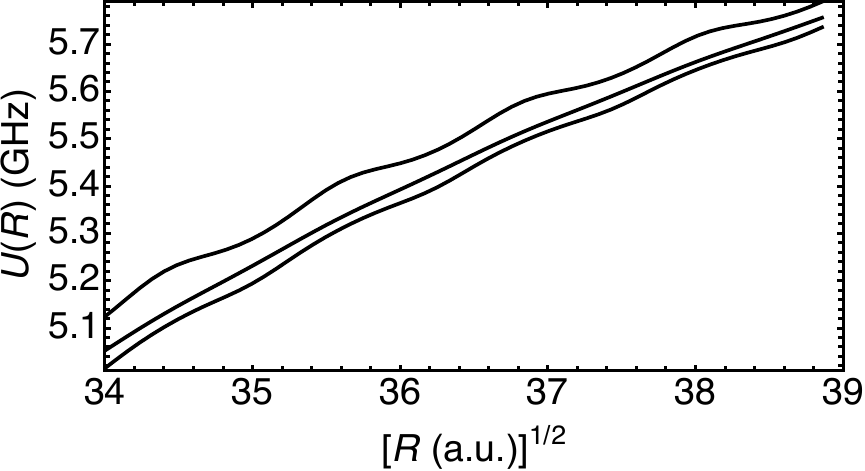}
		\caption{\label{fig:butterfly70}  Potential energy curves of the $M_\text{tot}=\frac{1}{2}$ singlet butterfly potential curves of Rb$^*$Rb, with the threshold energy set at the hydrogenic $n=70$ level with an $f=2$ perturber. The reference energy is set to the $f=1$ level. The $\Lambda$ doublet (lower two curves) has split due to the relatively strong hyperfine coupling. 
		}
	\end{figure}

The hyperfine splitting also plays a role in breaking the degeneracy of the butterfly potential energy curves in the singlet-dominated $P$-wave scattering channel. The inset of Fig.~\ref{fig:singlettotriplet23} highlights these levels.  
The oscillating potential curve is, to an excellent approximation, a $\Sigma$ molecular state having $M_L = 0$, while the smooth potential curve is actually two nearly degenerate $\Pi$ curves having $M_L = \pm 1$. 
For this $n$, these energy levels are split by approximately 1 Hz, making them challenging to obtain with the Green's function. 
Again, as a function of $n$, the role of the hyperfine coupling changes, and at high $n$ it plays a key role in breaking this degeneracy. The potential curves for $n = 70$ shown in Fig.~\ref{fig:butterfly70} demonstrate this, where now the $M_L$ states are coupled and the degeneracy is broken on the MHz scale. 
 
	\section{Discussion}
	\label{sec:discussion}

The level of spectroscopic accuracy currently attainable in Rydberg molecule experiments reaches the level of a few MHz. 
Although the dominant sources of inaccuracy in the calculation of potential energy curves are eliminated in our method, there are still corrections outside the scope of this theory which could hinder comparison of predicted and observed binding energies.

One source of error is our truncation to partial waves $L\le 1$, which is the standard assumption made in the literature, with the sole exception being Ref. \cite{Giannakeas2020b}). 
 As shown there, the main effect of including higher partial waves is to induce additional ``trilobite"-like potential curve which descend from the degenerate manifolds. 
These potential curves are proportional to $\sim -(RL^3)^{-1}$, and are therefore suppressed at large $R$. 
We estimate the contribution of $L=2$ scattering in Appendix \ref{app:highL}, concluding that it increases the depth of the potential wells by approximately one percent. 

A second source of error is non-adiabatic coupling, which is typically neglected in studies of long-range Rydberg molecule.
Only a few exceptions have considered these corrections \cite{hummelVibronic2023,UltracoldChem,srikumarNonadiabatic2023}. 
There is no inherent limitation in using the Green's function method to compute non-adiabatic effects, but the computation of the electronic eigenstates and their derivatives, the key ingredients of non-adiabatic coupling terms, is beyond the present scope of this paper. 
We have estimated the diagonal coupling term due to non-adiabatic physics in Appendix \ref{app:NAC}. It is inconsequential for vibrational levels bound in potential wells isolated from extremely sharp avoided crossings for most $n$ values, although at higher $n$ the shifts can become measurable in trilobite and butterfly states \cite{hummelVibronic2023}. 

    Another source of error derives from the fact that in electron scattering from an atom with no permanent electric quadrupole moment, the dominant long range interaction is the induced dipole polarizability, varying asymptotically like $r^{-4}$.  This produces a modification of the Wigner threshold law for all partial waves $L>0$ which in turn causes all non-$S$-wave generalized Fermi-Omont pseudopotentials to diverge at zero collision energy\cite{OMalley1961,Holzwarth1973,WatanabeGreene1980}.  (E.g., for $p$-wave collisions, the coefficient of the Omont contact potential is $-6 \pi \tan{\delta_P}/k(R)^3$ while $\delta_P \propto k(R)^2$ at $k(R) \rightarrow 0$.)  Ideally, those $L>0$ partial wave phaseshifts should be broken into a short-range contribution whose effects can be incorporated as a zero-range pseudopotential, plus a long-range contribution that causes the modified non-Wigner threshold law treated perturbatively as an explicit polarizability potential between the electron and the perturbing atom.  The development of this type of formulation could be important for high-precision applications, and remains a goal for future studies.
	Likewise, we rely on the extrapolation of phase shifts  at large $R$ when the semiclassical energy becomes negative.

	\section{Conclusions}
This article develops and implements a fully spin-dependent Green's function method to compute the adiabatic potential energy curves of long-range Rydberg molecules. 
Our method is directly applicable to any combination of alkali atoms and, with appropriate extensions, to alkaline earth atoms. 
Potential curves are obtained from this method by the numerical solution of the roots of a determinantal equation. 
Nearly all of the dependence on the Rydberg wave functions and energy is handled analytically, and therefore the dimensionality the matrix in this determinant does not increase with $\nu$, facilitating the study of very high principal quantum numbers. 

The discussion above emphasizes how the use of this more accurate approach, which eliminates the ambiguity related to the non-convergent diagonalization method that has plagued other studies, should enable more accurate studies of Rydberg molecule spectra and the electron-atom scattering phase shifts upon which they depend. 
Having more accurate calculations at hand should also improve the accuracy of proposals involving the exaggerated properties of Rydberg molecules, such as those to study the strong multipolar interactions between long-range Rydberg molecules \cite{EilesPendular,rivera-rodriguezElectrostatic2021a}, the formation of heavy Rydberg states out of long-range molecules \cite{peper2020formation,hummel2020ultracold}, and the behavior of Rydberg atoms within dense ultracold gases \cite{eilesAnderson,camargoCreation2018a,sous,hofferberth}. 
 
	\acknowledgements
            The work of C.H.G. has been supported in part by NSF grant No.2207977.  C.H.G. also appreciates discussions with Edward Hamilton and access to his unpublished Ph.D. thesis and related computer programs.
			M.T.E was supported by an Alexander von Humboldt Stiftung during the early stages of this research. 
			M.T.E. is grateful to have benefited from discussions with and help from P. Giannakeas, C. Fey, F. Hummel, A. Eisfeld, Č. Lozej, and A. A. T. Durst. 

   A python implementation of this method and the root-finding procedure is available upon reasonable request from \href{mailto:matt.eiles1@gmail.com}{matt.eiles1@gmail.com}.
	\appendix
	\section{Alternative expression for the Coulomb Green's function}
 For completeness, we state here 
	another useful expression for the Coulomb Green's function, which utilizes a partial wave expansion in terms of standard Whittaker functions,
	\begin{align}
		G^{C}(\vec{r},\vec{r} \, ' \! ;\nu) &=\sum_{\ell=0}^\infty  \frac{\nu \Gamma(\ell+1-\nu)}{2rr' (2\ell+1)!}\\&\nonumber\times M_{\nu,\ell+1/2}\left(\frac{2r_<}{\nu}\right) W_{\nu,\ell+1/2}\left(\frac{2r_>}{\nu}\right)\\&\times\nonumber \sum_{m_\ell} \sph_{\ell,m_\ell}^{*}({\hat r'})\sph_{\ell,m_\ell}({\hat r}).
	\end{align}
	
	\begin{widetext}

		\section{ Taylor expansions of $G^C$}
		\label{app:taylor}
		Our development requires some Taylor expansions of  $G^C(\vec r, \vec r\,')$  for $Y \ll X \ll 1$. 
		Many of these terms were first computed by E. L. Hamilton \cite{hamiltonPhotoionization}.  
		First of all, we introduce a convenient notation:
		\begin{align}
			v &= |\vec{r}-\vec{r} \, ' | = \nu \frac{( \xi - \eta )}{2}\\
			u &= r + r' = \nu \frac{( \xi + \eta )}{2}.
		\end{align}
		Then the following expansions will prove to be useful:
		\begin{align}
			\label{eq:u2R}
			u - 2R &= (X \cos \theta_{X} + Y \cos \theta_{Y}) + \mathcal{O}({1}/{R})\\ &\approx \frac{4\pi}{\sqrt{3}} \label{eq:u2R2}
			\left(  X \sph_{10}({\hat X}) \sph_{00}^*({\hat Y})+ Y \sph_{00}(\hat X)\sph_{10}^*(\hat Y) \right) \\
			\label{eq:v}
			v &= \sqrt{X^{2} + Y^{2} - 2X Y\cos\gamma},\\
			\text{where }\cos\gamma &=\sin \theta_{X} \sin \theta_{Y}
			\cos(\phi_{X} - \phi_{Y}) + \cos \theta_{X} \cos \theta_{Y})
		\end{align}
		Eq. \ref{eq:u2R2} is expressed in terms of spherical harmonics $\sph_{LM}(\hat X)$ with $L <2$, and in the following expansions we will continue to write the expansions in these same terms. 
		As we include only $S$ and $P$ partial waves in the expansion in Eq. \ref{eq:wfnearperturber1}, any higher order terms with $L\ge 2$ will have vanishing matrix elements. 
		Our Taylor expansion of $G^C$ will initially be represented in powers of $(u-2R)$ and $v$. 
		An alternative expansion for $v=|\vec{X}-\vec{Y}|$ in terms of spherical harmonics is \cite{Varsh}
		\begin{equation}
			v=4\pi \sum_L  \frac{1}{2L+1} \frac{Y^L}{X^{L+1}} \left(\frac{Y^2}{2L+3}
			-\frac{X^2}{2L-1} \right) 
			\sum_M \sph_{LM}(\hat X)\sph_{LM}^*(\hat Y),
		\end{equation}
		which, truncated to $L< 2$ is
		\begin{eqnarray}
			v\approx 4\pi \left(X+\frac{Y^2}{3X}\right)\sph_{00}(\hat X)\sph_{00}^*(\hat Y)-
			\frac{ 4\pi}{3} \left(Y-\frac{Y^3}{5 X^2}\right) \sum_M \sph_{1M}(\hat X)\sph_{1M}^*(\hat Y).
		\end{eqnarray}
		Squaring Eqs.~\ref{eq:u2R} and \ref{eq:v} yields
		\begin{align}
			v^2 &= X^2+Y^2-2 X Y \frac{4\pi}{3} \sum_M \sph_{1M}(\hat X)\sph_{1M}^*(\hat Y),\\
			(u-2R)^2 &\approx X^2 \cos^2\theta_X + Y^2 \cos^2\theta_Y + 2 X Y \cos\theta_X \cos\theta_Y.
		\end{align}
		Retaining terms at most quadratic and of spherical harmonic orders $L\le 1$, we can write:
		\begin{eqnarray}
			(u-2R)^2 \approx \frac{4\pi}{3} \left[ (X^2+Y^2)  \sph_{00}(\hat X)\sph_{00}^*(\hat Y) + 2 X Y \sph_{10}(\hat X)\sph_{10}^*(\hat Y) \right].
		\end{eqnarray}
		Lastly, one mixed expansion is required, where only terms that are at most linear in each  variable are retained. In terms of spherical harmonics, this is
		\begin{eqnarray}
			(u-2R)v \approx  X^2 \frac{4 \pi}{\sqrt{3}} \sph_{10}(\hat X)\sph_{00}^*(\hat Y)+X Y 
			\left(\frac{8 \pi}{3\sqrt{3}}
			\sph_{00}(\hat X)\sph_{10}^*(\hat Y) + 
			{\rm \ terms\ of\ rank\ }L=2.\right)
		\end{eqnarray}
		With these preliminaries out of the way, we evaluate the Taylor series of $G^C(u,v;\nu)$ for small values of $u-2R$ and $v$:
		\begin{align}
			2\pi G^C(u, v; \nu) & = \frac{1}{ v} \Phi(u, v; \nu) \nonumber\\
			& \approx   \frac{1}{v} + \Phi_{v} + \Phi_{uv} (u-2R)  + \frac{1}{2} \Phi_{vv} v
			+ \frac{1}{2} \Phi_{uuv} (u-2R)^{2} \nonumber 
			+ \frac{1}{2} \Phi_{uvv} (u-2R)v + \frac{1}{6} \Phi_{vvv} v^{2}  . 
		\end{align}
		The various $\Phi$ terms with subscripts here denote derivatives of the Green function with respect to $u$ or $v$ and are given by 
		\begin{align}
			\label{eq:DerivList1}
			\Phi_{vv} & = - k^{2} \\
						\label{eq:DerivList2}
			\Phi_{uv} & = - \frac{\nu \Gamma(1-\nu)}{2R^{2}} M_{\nu} W_{\nu} \\
						\label{eq:DerivList3}
			\Phi_{vvv} & =  \Gamma(1-\nu) \left\{ \left(\frac{\nu}{2R^{3}} -
			\frac{k^{4}\nu}{2}\right) M_{\nu} W_{\nu} -
			\frac{2k^{2}}{\nu} M_{\nu}' W_{\nu}' + \frac{1}{2R^{2}} \left( M_{\nu}' W_{\nu} + M_{\nu} W_{\nu}' \right)		
			\right\}  \\
						\label{eq:DerivList4}
			\Phi_{uvv} & =  \frac{2}{R^{2}} \\
						\label{eq:DerivList5}
			\Phi_{uuv} & = \Gamma(1-\nu) \left\{ \frac{\nu}{2R^{3}} M_{\nu} W_{\nu} -
			\frac{1}{2R^{2}} \left(M_{\nu}' W_{\nu} + M_{\nu} W_{\nu}'\right)
			\right\}\\
						\label{eq:DerivList6}
			\Phi_{v}  &=  \frac{\Gamma (1-\nu )}{2\nu R^2} M_{\nu }\left(\left(\nu ^2
			\left(R^2+1\right)-R^2\right) W_{\nu}+\nu  R (\nu-\nu R +R-1) W_{\nu ,\frac{3}{2}}\right)
			\\
			&+\frac{\Gamma (1-\nu ) }{12
				\nu R}\left(\nu ^2-1\right) M_{\nu ,\frac{3}{2}} \left(\nu  (R-1) W_{\nu}-(\nu -1) R W_{\nu
				,\frac{3}{2}}\right).
		\end{align}
		For brevity, we omit the second subscript on the Whittaker functions when it is equal to 1/2 and do not write out the full argument $2R/\nu$, i.e. $M_\nu \equiv M_{\nu,\frac{1}{2}}(\frac{2R}{\nu})$, etc. 
		Primes denote derivatives with respect to the full argument. 
		Finally, we insert all expansions into this expression and regroup the terms
		according to their angular
		character relative to the perturber center:
		\begin{align}
			\label{eq:TaylorExpRegroup}
			G^{C}(u, v; \nu) & =  \frac{1}{2 \pi} \biggr\{ \left(\frac{1}{X} + \Phi_{v} +
			\frac{1}{2} \Phi_{vv} X + \frac{1}{6} \Phi_{vvv} X^{2}+\frac{1}{3}\Phi_{uuv} X^{2}) \right. {4\pi} \sph_{00}(\hat X)\sph_{00}^*(\hat Y) \\
			&  + \left(\Phi_{uv} X + \frac{1}{2} \Phi_{uvv} X^{2} \right) \frac{4\pi}{\sqrt{3}} \sph_{10}(\hat X)\sph_{00}^*(\hat Y) \nonumber \\
			&  + \left(\Phi_{uv} Y + \frac{1}{3} \Phi_{uvv} XY \right) \frac{4\pi}{\sqrt{3}} \sph_{00}(\hat X)\sph_{10}^*(\hat Y) + \Phi_{uuv}XY  \frac{4\pi}{3} \sph_{10}(\hat X)\sph_{10}^*(\hat Y)   \nonumber \\
			&  + \left(\frac{Y}{X^{2}} - \frac{1}{2} \Phi_{vv} Y -
			\frac{1}{3} \Phi_{vvv} XY \right)  \frac{4\pi}{3}  \sum_{M=-1}^1 \sph_{1M}(\hat X)\sph_{1M}^*(\hat Y)
			\biggr\}. \nonumber
		\end{align}
		As discussed in the text, the eventual matrix representation of the solution requires integrating the Green's function and its derivative over the $\hat X$ and $\hat Y$:
		\begin{align*}
			\Delta^{M_L}_{L,L'}(X,Y)  &= \oint\text{d}\hat X\oint \text{d}\hat Y \sph_{LM_L}^*(\hat X) G^C_\nu(\vec{X},\vec{Y})\sph_{L'M_L}(\hat Y),  \\
			\bar{\Delta}^{M_L}_{L,L'}(X,Y)  &= \oint\text{d}\hat X\oint \text{d}\hat Y \sph_{LM_L}^*(\hat X)\partial_Y G^C_\nu(\vec{X},\vec{Y})\sph_{L'M_L}(\hat Y).
		\end{align*}
		We evaluate these integrals analytically in the vicinity of the perturber by using only the relevant terms of Eq. \ref{eq:TaylorExpRegroup}, dropping higher order terms in $X$ and $Y$. 
		In a matrix notation in the space $L = \{0,1\}$ these read
		\begin{align}
			\label{eq:del}
			\Delta_{L,L^{\prime }}^{(M_{L^{\prime }}=0)}&=2\begin{pmatrix}
				\frac{1}{X}+\Phi _{v} & \frac{1}{\sqrt{3}}\Phi _{uv}Y \\
				\frac{1}{\sqrt{3}}\Phi _{uv}X & \frac{1}{3}[
				\frac{Y}{X^{2}}-\frac{1}{2}\Phi _{vv}Y-(\frac{1}{3}\Phi _{vvv}-\Phi
				_{uuv})XY]
			\end{pmatrix}\\
			\label{eq:delbar}
			\bar{\Delta}_{L,L^{\prime }}^{(M_{L^{\prime }}=0)}&=2\begin{pmatrix}
				0 & \frac{1}{\sqrt{3}}\Phi _{uv} \\
				0 & \frac{1}{3}[\frac{1}{X^{2}}-\frac{1}{2}\Phi _{vv}-(\frac{1}{3}\Phi
				_{vvv}-\Phi _{uuv})X]\end{pmatrix}\\
			\Delta_{1,1}^{M_L'=\pm 1} &= \frac{2Y}{3}\left(\frac{1}{X^2} - \frac{1}{2}\Phi_{vv} - \frac{1}{3}\Phi_{vvv}X\right).\label{eq:dell}
		\end{align}
		Note that the lack of symmetry in these matrices stems from the asymmetry in the $X$ and $Y$ variables.

		\section{Taylor expansion of the Whittaker function}
		\label{app:taylorwhit}
		In this appendix, we develop the Taylor expansion of the Whittaker function multiplied by an ion-centered spherical harmonic in the vicinity of the perturber, i.e. 
		\begin{align}
			F(\vec{r})\equiv\frac{1}{r}{\cal W}_{\nu_f,\ell +1/2}(r) \sph_{\ell m_\ell}(\hat r)\approx F(\vec{R})+\vec{X}\cdot(\nabla_R F(\vec{R})). 
		\end{align}
		Evaluating this explicitly,
		\begin{align}
			F(\vec r)\approx \frac{1}{R}{\cal W}_{\nu_f,\ell +1/2}(R) \sph_{\ell m_\ell}(0,0)+X \cos{\theta_X}\partial_R \frac{{\cal W}_{\nu_f,\ell +1/2}(R)}{R}Y_{\ell m_\ell}(0,0)-i\frac{{\cal W}_{\nu_f,\ell +1/2}(R)}{R^2}\vec{X}\cdot {\hat r} \times \vec{L}\sph_{\ell m_\ell}(\hat r)|_{0,0}.
		\end{align}
		Using
		\begin{equation}
			\vec{X}\cdot {\hat r} \times \vec{L} =-X \sin{\theta_X} \cos{\phi_X}\frac{L_+-L_-}{2i}+X\sin{\theta_X} \sin{\phi_X} \frac{L_+ + L_-}{2}
		\end{equation}
		gives a fairly simple final expression,
		\begin{align}
			-i\vec{X}\cdot {\hat r} \times\vec{L}\sph_{\ell m_\ell}(\hat r)|_{0,0}=
			\frac{1}{2}X \sin{\theta_X} \Bigg( &e^{-i \phi_X} \sph_{\ell m_\ell+1}(0,0)\sqrt{\ell(\ell+1)-m_\ell(m_\ell+1)}\\& -  e^{i \phi_X} \sph_{\ell m_\ell-1}(0,0)\sqrt{\ell(\ell+1)-m_\ell(m_\ell-1)}  \Bigg),
		\end{align}
		which evidently vanishes unless $m_\ell = \pm 1$. Evaluating the right hand side for these two $m_l$ values yields
		\begin{equation}
			-i\vec{X}\cdot {\hat r} \times\vec{L}\sph_{\ell m_\ell}(\hat r)|_{0,0} = X \sqrt{\frac{\ell(\ell+1)(2\ell+1)}{6}} {\sph^*_{1 -m_\ell}({\hat X})} \delta_{m_\ell,\pm 1}.
		\end{equation}
		The final equations can be written more simply after introducing a more compact notation.  We define the coefficients $b^{(\ell,\nu_f)}_{L M_L}$ as follows:
		\begin{align}
			\label{eq:b00}
			b_{0,0}^{(\ell,\nu_f)}&=\frac{\sqrt{2\ell+1}}{R} {\cal W}_{\nu_f,\ell+1/2}(R)  \\
			\label{eq:b10}
			b_{1,0}^{(\ell,\nu_f)}&=\sqrt{\frac{2\ell+1}{3}} \partial_R \frac{{\cal W}_{\nu_f,\ell+1/2}(R)}{R}  \\
			\label{eq:b11}
			b_{1,\pm 1}^{(\ell,\nu_f)}&=\sqrt{\frac{\ell(\ell+1)(2\ell+1)}{6}} \frac{{\cal W}_{\nu_f,\ell+1/2}(R)}{R^2}.
		\end{align}
		These allow the first-order Taylor expansion to be written as
		\begin{equation}
			\frac{1}{r} {\cal W}_{\nu_f,\ell+1/2}(r) \sph_{\ell M_L}({\hat r}) \equiv \sum_{L=|M_L|}^1 X^L b^{(\ell,\nu_f)}_{L M_L} \sph_{LM_L}({\hat X}). 
		\end{equation}
		To avoid differentiating the (ordinary) Whittaker function, the  following identity can be used:
		\begin{equation}
			\partial_R \frac{{\cal W}_{\nu_f,\ell+1/2}(R)}{R} =  \frac{\left[ R-\nu(\nu+1)\right] W_{\nu,\ell+1/2}(\frac{2R}{\nu})-\nu W_{\nu+1,\ell+1/2}(\frac{2R}{\nu})}{R^2[\nu \Gamma(\nu+\ell+1)\Gamma(\nu-\ell)]^{1/2}}   .\end{equation}

		\section{Green's function formulation with no spins or quantum defects}
		In this appendix we derive the potential energy curves in the absence of spins and quantum defects. This clarifies the more complicated spin-dependent derivation in the main text and allows for comparisons with the generalized local frame transformation theory method of Ref. \cite{Giannakeas2020a} and the Green's function results of Ref. \cite{hamiltonPhotoionization}. 
		
		We begin with the following integral equation for the electronic wave function, valid everywhere outside of a small volume of radius $Y$ around the perturber:
		\begin{equation}
			\Psi(\vec{r}) = \frac{1}{2} \oint \left\{
			\frac{\partial G^{C}(\vec{r},\vec{r} \, ')}{\partial Y}
			\Psi(\vec{r} \, ')
			-G^{C}(\vec{r},\vec{r} \, ')\frac{\partial \Psi(\vec{r} \, ')}{\partial Y} 
			\right\} da \, '.
			\label{eq:OuterSolution3app}
		\end{equation}
		The integrals in this expression are intended to be taken over the surface of a small sphere of radius $Y$ centered on the perturber, and in fact we will simplify many of our final expressions by taking the limit $Y \rightarrow 0$.  The integrals involve not only the Green's function, but also the wavefunction $\Psi$ and its radial derivative near the perturber.
		We describe the electronic wave function near the perturber using the partial wave expansion 
		\begin{equation}
			\label{eq:appGF1}
			\langle Y|\Psi \rangle = \sum_{L=|M_L|}^1 |L,M_L \rangle \left[ j_L(k Y) \cos{\delta^L}-y_L(k Y) \sin{\delta^L} \right]B_L,
		\end{equation}
		The semiclassical de Broglie wavenumber for the Rydberg electron at the point of collision with the perturber defines  $k=\sqrt{-\frac{1}{\nu^2}+\frac{2}{R}}$.
		To turn the integral equation of Eq. \ref{eq:OuterSolution3} into a matrix equation, we insert Eq. \ref{eq:appGF1} into \ref{eq:OuterSolution3} and project from the left over angles $\hat X$ at a small value of $X$, although constraining $X>Y$. We can write our equation symbolically as:
		\begin{equation}
			_X\langle L',M_L | \Psi \rangle = \frac{ Y^2}{2}  \left( _X\langle L',M_L | \partial_YG^C_\nu(\vec{X},\vec{Y}) |\Psi \rangle_Y -
			_X\langle L',M_L| G^C_\nu(\vec{X},\vec{Y}) |\partial_Y\Psi \rangle_Y  \right).
		\end{equation}
		Here the implied integrals are taken over $d\Omega_Y$ and $d\Omega_X$.  This is made more explicit by plugging in the expansions for $\Psi$ very close to the perturber:
		\begin{align}
			\label{eq:appGF2}
			\left( j_{L'}(k X) \cos{\delta^{L'}}-y_{L'}(k X) \sin{\delta^{L'}} \right)B_{L'} &= \nonumber  
			(Y^2/2)\sum_L {\biggr (} \overline{\Delta}_{L,L'}^{M_L}(X,Y) \left( j_L(k Y) \cos{\delta^L}-y_L(k Y) \sin{\delta^L} \right)  \nonumber \\
			&-\Delta_{L,L'}^{M_L}(X,Y) \partial_Y \left( j_L(k Y) \cos{\delta^L}-y_L(k Y) \sin{\delta^L} \right) {\biggr )}  B_L,
		\end{align}
		where $\Delta_{L,L'}^{M_L}$ and $\overline{\Delta}_{L,L'}^{M_L}$ are defined in Eqs. \ref{eq:aafb} and \ref{eq:del}-\ref{eq:dell}.
		Equation \ref{eq:appGF2}, for any chosen surfaces (assumed to be small, with $X>Y$), can be cast as a determinantal equation whose eigenroots $\nu$ will be the electronic energies at a given value of $R$, i.e. the Born-Oppenheimer potential curves. 
		We require a handful of expansions of the spherical Bessel functions, and retain only the first few terms for $kX,kY\ll 1$. Using these, we evaluate Eq. \ref{eq:appGF2}, keeping the lowest order non-vanishing terms in the expansion. 
		This yields
		\begin{align}
			0&= -B_0 (k^2 \cos{\delta_0}-\Phi_v k \sin{\delta_0})+ B_1 \sqrt{3}\Phi_{uv}\sin{\delta_1} \\
			0&= -B_1[k^3 \cos{\delta_1}+(\Phi_{vvv}-3\Phi_{uuv})\sin{\delta_1}]+\sqrt{3} B_0 \Phi_{uv} k \sin{\delta_0}.
		\end{align}
		These two homogeneous equations imply a transcendental equation
		\begin{equation}
			0 = \left(1 - \Phi_v\frac{\tan\delta_0}{k}\right)\left[1+\left(\Phi_{vvv} - 3\Phi_{uuv}\right)\frac{\tan\delta_1}{k^3}-\frac{\tan\delta_0/k}{1 - \Phi_v\tan\delta_0/k}3\Phi_{uv}^2\frac{\tan\delta_1}{k^3}\right]
		\end{equation}
		This agrees with Hamilton's transcendental equation (4.30) for $\Sigma$ states when quantum defects vanish, correcting a typo in the progression from Eq. 4.29. The factorization shows the trilobite and butterfly terms emerging naturally, along with the (typically weak) coupling between them in the third term inside the brackets.   The corresponding equation for $\Pi$ states is
		\begin{equation}
			B_1\left(1+\Phi_{vvv}\frac{\tan\delta_1}{k^3}\right)=0.
		\end{equation}
		
		\section{Recoupling matrix element}
		\label{app:recoup}
		The key recoupling quantity, 
		\begin{align}
			\mathcal{A}_{i\alpha}\equiv\langle  {\bf i} | \boldsymbol{\alpha}\rangle &=\bkt{L_iM_{L_i} s_R m_{R_i}, (Is_{p})f m_{f} }{(SL_\alpha)J M_{J},I m_{I_\alpha}} \\
			&=\sum_{m_{R_\alpha},m_{p_\alpha},M_{S_\alpha},M_{L_\alpha}}
			\sum_{m_{p_i},m_{I_{i}}} C_{S M_{S},L_\alpha M_{L_\alpha}}^{J M_{J}} C_{s_R m_{R_\alpha},s_pm_{p_\alpha}}^{S M_{S}} C_{s_p m_{p_i},I m_{I_i}}^{f m_{f}}\nonumber \\ &\times \delta_{{L_\alpha},{L_i}} \delta_{M_{L_\alpha},M_{L_i}} \delta_{m_{R_\alpha},m_{R_i}} \delta_{m_{p_\alpha},m_{p_i}}   \delta_{m_{I_\alpha},m_{I_i}}, \nonumber
		\end{align}
		where the Kronecker delta functions in the last line allow us to eliminate four out of the six sums. This gives the following simplified expression:
		\begin{align}
			\label{eq:recoupApp}
			\mathcal{A}_{i\alpha}=
			\delta_{{L_\alpha},{L_i}}\sum_{M_{S},{m_{p}}}
			C_{S M_{S},L_\alpha M_{L}}^{J M_{J}} C_{s_R m_{R},s_pm_{p}}^{S M_{S}} C_{s_p m_{p},I m_{I}}^{f m_{f}}.
		\end{align}
  \end{widetext}

  \section{Corrections to the PECs}
This Appendix considers the main sources of error outside the current scope of our theory and estimates their strengths using a minimal diagonalization calculation. 
We consider the two Rydberg states studied in detail in the main text and focus on the adiabatic potential curve connecting the asymptotic Rydberg $(n+2)p$ states with the butterfly state at low $R$. This is denoted the ``$p$-butterfly" PEC. 
The Rydberg basis is truncated to include the Rydberg manifold in question and the one below in order to stabilize the $P$-wave shape resonance. 
We ignore all spin degrees of freedom and use the Rb fine structure-averaged phase shifts $\mu_s = 3.13$, $\mu_p=2.642$, $\mu_d= 1.348$, and $\mu_f= 0.017$. 
Here only triplet scattering is included, using the $J$-independent phase shifts calculated by Ref. \cite{eilesFormation2018}. 
The $p$-butterfly PEC is shown for both $n$ levels considered here in the upper panel of Figs. \ref{fig:23corrections},\ref{fig:14corrections}. 
Its depth ranges, as $n$ decreases from 23 to 14, from a few hundred MHz to a few GHz in the $p$-state regime (see the inset of Fig. \ref{fig:23corrections}) and a few tens of GHz to a few hundred GHz in the butterfly regime. 

  \subsection{Higher-order partial wave scattering}
  \label{app:highL}

 		\begin{figure}[t]
		\includegraphics[width=0.9\columnwidth]{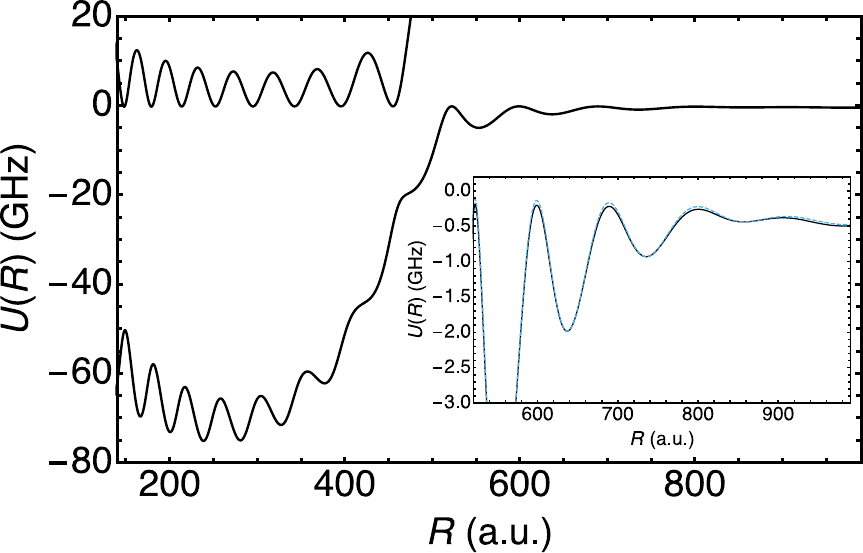}\\
  \includegraphics[width=0.9\columnwidth]{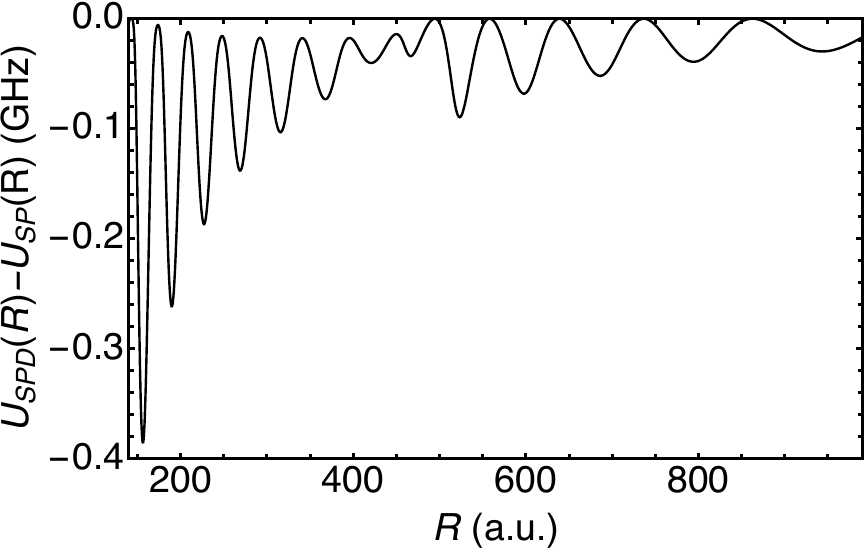}
		\caption{\label{fig:23corrections}  
 The upper panel shows the $p$-butterfly potential energy curve, with threshold value set to the $25p$ asymptotic energy, and the lower panel shows the difference between this potential curve $U_{SP}(R)\equiv U(R)$ and the one, $U_{SPD}(R)$, including $D$-wave physics. }
	\end{figure}
 
\begin{figure}[t]
		\includegraphics[width=0.9\columnwidth]{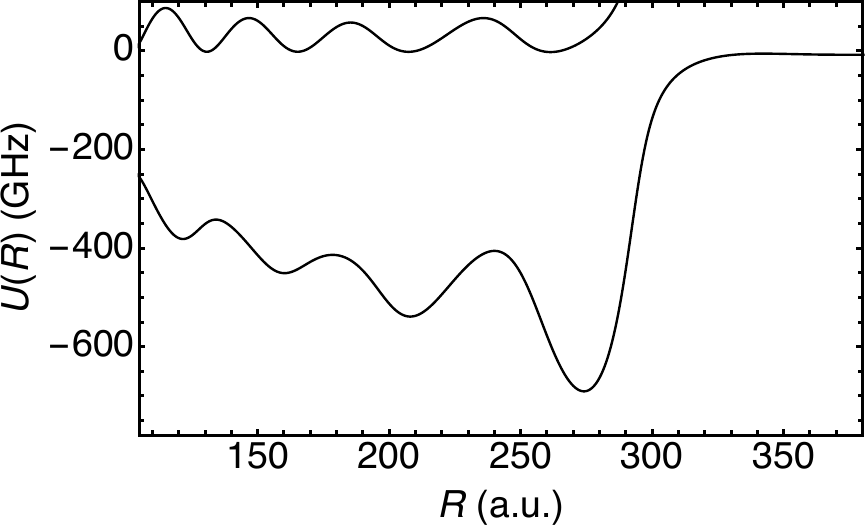}\\
  \includegraphics[width=0.9\columnwidth]{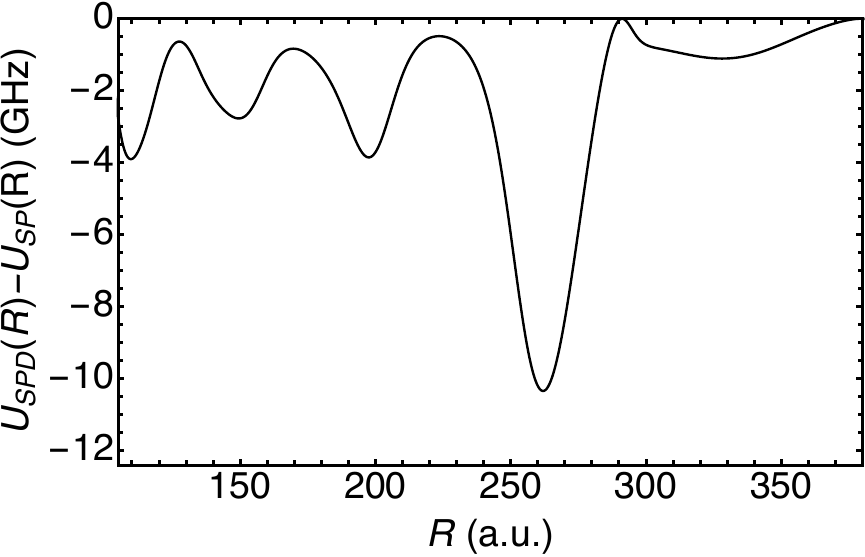}
		\caption{\label{fig:14corrections}  
The upper panel shows the $p$-butterfly potential energy curve, with threshold value set to the $16p$ asymptotic energy, and the lower panel shows the difference between this potential curve $U_{SP}(R)\equiv U(R)$ and the one, $U_{SPD}(R)$, including $D$-wave physics. }
	\end{figure}
Next, consider the effect of $D$-wave interactions on the other potential curves using this minimal model. 
The matrix elements of the $D$-wave pseudopotential are evaluated explicitly in the supplementary information of Ref.~\cite{Giannakeas2020b}, where the $D$-wave phase shifts can also be found. 
For $L>1$ the phase shifts are well-described by the partial wave Born approximation, which gives $\delta_L(k) = \frac{\pi \alpha k^2}{(4L^2-1)(2L+3)} $.
   The lower panel of Fig. \ref{fig:23corrections} shows the difference between the $p$-butterfly PEC computed with ($U_{SPD}(R)$) and without ($U_{SP}(R)$) the $D$-wave interaction. The correction is on the order of about one percent, increasing which leads to shifts in the vibrational energies of $\sim60-80$ MHz in the butterfly region (where the overall bound states are $\sim 50-75$ GHz deep) and shifts of $\sim 10-30$ MHz in the long-range $P$-state region, where the binding energies are a few hundred MHz.

   At smaller $n$ the absolute strength of the $D$-wave interaction is larger, as seen in Fig. \ref{fig:14corrections}, but the relative strength remains about one percent. The vibrational bound states are shifted by about 6GHz out of 600GHz in the butterfly region here as well. 
  \subsection{Non-adiabatic corrections}
    \label{app:NAC}

Non-adiabatic physics modifies results from the Born-Oppenheimer picture in two ways: it couples adiabatic potentials together through the first-derivative coupling term often called the ``$P$-matrix", and it adds an overall repulsive correction term due to the second-derivative coupling matrix. 
This correction, often called the ``Born-Huang" term, is  
\begin{equation}
    U_{NAD}(R) = -\frac{1}{2\mu}\left\langle\Psi_U( R)
    \left|\frac{\partial^2}{\partial R^2}\right|\Psi_U( R)\right\rangle_r
\end{equation}
Here, $\ket{\Psi_U(R)}$ denotes the adiabatic eigenstate corresponding to the potential energy curve $U(R)$.

This correction term is computed for both of the $n=14$ and $n=23$ $p$-butterfly PECs, again using our minimal diagonalization model. 
As shown in  Fig. \ref{fig:14corrections}, this correction is always positive and its size is nearly independent of $n$. 
For the $p$-butterfly calculations considered in the main text, this term is sufficiently small to neglect, but its importance should be re-considered for higher $n$ levels where the depth of the potentials decreases to the few GHz level \cite{hummelVibronic2023}. 
 	
  \begin{figure}[t]
		\includegraphics[width=0.9\columnwidth]{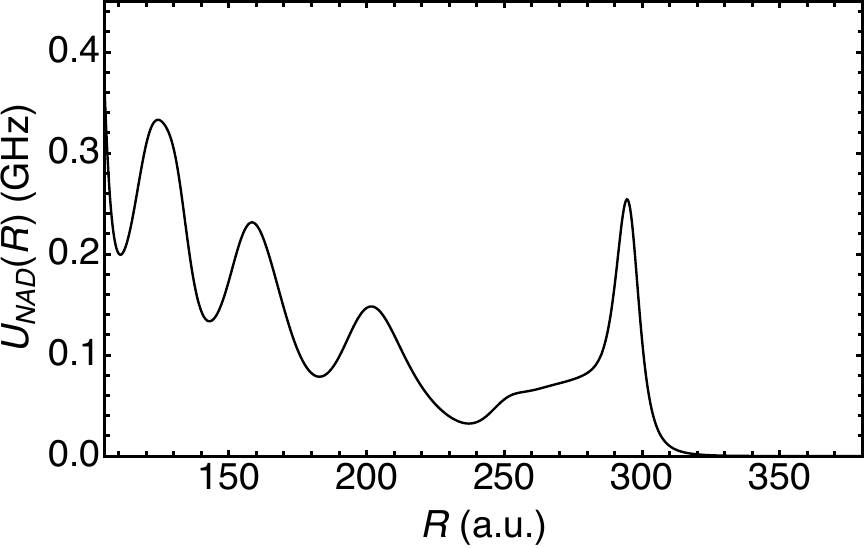}\\
  \includegraphics[width=0.9\columnwidth]{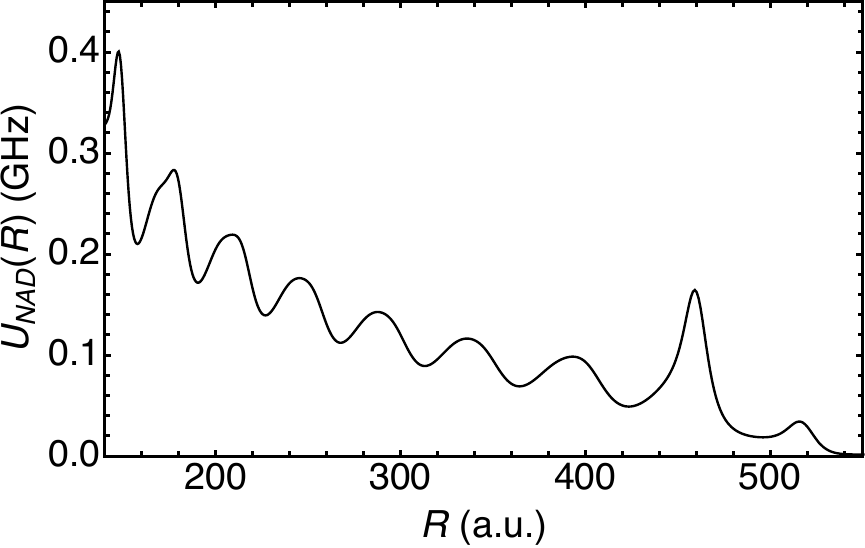}
		\caption{\label{fig:nadcorrections}  
The Born-Huang correction for $16p$-butterfly  (upper panel) and $23p$-butterfly (lower panel) potential energy curves. The correction term vanishes for larger $R$ in both cases as the state becomes almost completely a pure $p$ electronic state.  }
	\end{figure}

\bibliography{GFtreatment}
\end{document}